\documentclass[twocolumn,showpacs,preprintnumbers,amsmath,amssymb]{revtex4}
\usepackage{dcolumn}
\usepackage{bm}
\usepackage{epsfig}
\usepackage[usenames]{color}
\usepackage{hyperref}

\begin{document}
\title{An Analytical Window into the World of Ultracold Atoms}

\author{R. Radha}
\email{vittal.cnls@gmail.com}
\author{P. S. Vinayagam}
\affiliation{Centre for Nonlinear Science, PG and Research
Department of Physics, Government College for Women (Autonomous),
Kumbakonam 612001, India.}

\begin{abstract}
In this paper, we review the recent developments which had taken
place in the domain of quasi one dimensional Bose-Einstein
Condensates (BECs) from the viewpoint of integrability. To start
with, we consider the dynamics of scalar BECs in a time
independent harmonic trap and observe that the scattering length
can be suitably manipulated either to compress the bright solitons
to attain peak matter wave density without causing their explosion
or to broaden the width of the condensates without diluting them.
When the harmonic trap frequency becomes time dependent, we notice
that one can stabilize the condensates in the confining domain
while the density of the condensates continue to increase in the
expulsive region. We also observe that the trap frequency and the
temporal scattering length can be manoeuvred to generate matter
wave interference patterns indicating the coherent nature of the
atoms in the condensates. We also notice that a small repulsive
three body interaction when reinforced with attractive binary
interaction can extend the region of stability of the condensates
in the quasi-one dimensional regime.

On the other hand, the investigation of two component BECs in a
time dependent harmonic trap suggests that it is possible to
switch matter wave energy from one mode to the other confirming
the fact that vector BECs are long lived compared to scalar BECs.
The Feshbach resonance management of vector BECs indicates that
the two component BECs in a time dependent harmonic trap are more
stable compared to the condensates in a time independent trap. The
introduction of weak (linear) time dependent Rabi coupling rapidly
compresses the bright solitons which however can be again
stabilized through Feshbach resonance or by finetuning the Rabi
coupling while the spatial coupling of vector BECs introduces a
phase difference between the condensates which subsequently can be
exploited to generate interference pattern in the bright or dark
solitons.
\end{abstract}

\maketitle
\section{Introduction}
\begin{figure}
\begin{center}
\includegraphics[width=0.8\linewidth]{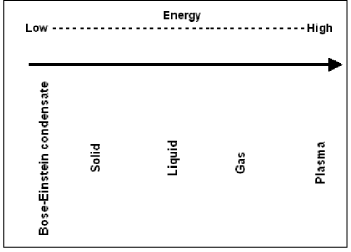}
\caption{The energy levels of different physical states of
matter.} \label{energydiagram}
\end{center}
\end{figure}

We are aware that eventhough matter pervades the entire universe,
it is found in just a few admissible forms such as solid, liquid
and gas. It is obvious that one can initiate a phase transition
between different states of matter by either increasing the
temperature or pressure. This understanding of generating new
states of matter by increasing the temperature was exploited in
1879 by Sir William Crookes \cite{crook79} to create ``plasma,'' a
gas containing non negligible number of charge carriers. It must
be mentioned that the physical states of matter change in going
from one phase to another while the chemical compositions of
matter remains the same. Can one go down the temperature scale and
generate a new state of matter at ultra cold temperatures? This
question has been plagueing the minds of scientists and it was in
1995, Cornell and Wieman \cite{ander95} created a ``Bose-Einstein
condensate'' (BEC) at super low temperatures. Eventhough
envisioned first by Albert Einstein and  a young Indian physicist
named Satyendra Nath Bose in the 1920s \cite{einst-einst,bose24},
it took more than seven decades to realize this singular state of
matter. In contrast to plasmas containing superhot and super
excited atoms, BECs were created at colder and colder temperatures
near absolute zero and they are composed of supercold and super
unexcited atoms (see Fig. (\ref{energydiagram})).
At such low temperatures, a large fraction of the atoms get piled
up either in the ground state or in the long lived metastable
state. In other words, the atoms merge together losing their
individual identities and behave like a giant matter wave. This
phenomenon is known as ``Bose-Einstein condensation.''

\section{Gross-Pitaevskii (GP) equation}
To investigate the dynamics of a Bose-Einstein condensate, we now
employ a Hartree or mean field approach and assume that the wave
function is a symmetrized product of single particle wave
functions. In the fully condensed state, all bosons (atoms with
integral spin) are in the same single-particle state,
$\phi(\bf{r})$ and therefore one can write down the wave function
of the $N$-particle system as
\begin{equation}
\Psi(\bf{r}_1, \bf{r}_2,\ldots,\bf{r}_N) = \prod_{i=1}^{N}
\phi(\bf{r}_i).\label{eq1.5}
\end{equation}

The single-particle wave function $\phi({\bf r}_i)$ is normalized
as
\begin{equation}
\int d{\bf{r}}|\phi({\bf{r}})|^2 =1.
\end{equation}

This wave function does not contain the correlations produced by
the interaction when two atoms are close to each other. These
effects are taken into account by using the effective interaction
$U_0 \delta({\bf r}-{\bf r'})$. According to mean field theory,
the effective Hamiltonian may be written as
\begin{equation}
H = \sum_{i=1}^{N}\left[\frac{{\bf{p}}_i^2}{2
m}+V({\bf{r}}_i)\right]+U_0
\sum_{i<j}\delta({\bf{r}}_i-{\bf{r}}_j),
\end{equation}
$V({\bf{r}}_i)$ being the external potential. The energy of the
state given by equation (\ref{eq1.5}) is written as
\begin{equation}
E = N\int d{\bf r}\left[\frac{\hbar^2}{2m}|\nabla\phi({\bf
r})|^2+V({\bf r})|\phi({\bf r})|^2+\frac{N-1}{2}U_0 |\phi({\bf
r})|^4\right].
\end{equation}

From the macroscopic theory for the uniform Bose gas, the relative
reduction of the number of particles in the condensate is of the
order of $(na^3)^{1/2}$, where $n$ is the particle density. If we
introduce the wave function of the condensed state as
\begin{equation}
\psi({\bf r})= N^{1/2} \phi({\bf r}),
\end{equation}
then the density of particles is given by
\begin{equation}
n({\bf r})= |\psi({\bf r})|^2
\end{equation}
and for $N\gg1$, the energy of the system may therefore be written
as

\begin{equation}
E = \int d {\bf r} \left[\frac{\hbar^2}{2m}|\nabla\psi({\bf
r})|^2+V({\bf r})|\psi({\bf r})|^2+\frac{1}{2}U_0 |\psi({\bf
r})|^4\right].
\end{equation} To find the optimal form for $\psi$, we minimize the energy with
respect to independent variations of $\psi(\bf{r})$ and its
complex conjugate $\psi^*(\bf{r})$, subject to the condition that
the total number of particles
\begin{equation}
N=\int d{\bf r} |\psi({\bf r})|^2
\end{equation}
be constant. For this, one writes $\delta E - \mu \delta N=0$
where the chemical potential $\mu$ is the Lagrange multiplier that
ensures constancy of the particles and the variations of $\psi$
and $\psi^*$ may thus be taken to be arbitrary. Equating to zero
the variation of $E-\mu N$ with respect to $\psi^*(\bf{r})$ gives
the following evolution equation \cite{gross61,gross63,pitae61}
\begin{equation}
-\frac{\hbar^2}{2m}\nabla^2 \psi({\bf r})+V({\bf r})\psi({\bf
r})+U_0 |\psi({\bf r})|^2\psi({\bf r})=\mu\psi({\bf
r}).\label{eq1.11}
\end{equation}

We call equation (\ref{eq1.11}) as the ``time-independent
Gross-Pitaevskii equation.'' This has the form of a
time-independent Schr\"{o}dinger equation in which the potential
acting on particles is the sum of the external potential $V({\bf
r})$ and a nonlinear term $U_0|\psi({\bf{r}})|^2$ that takes into
account the mean field produced by the other bosons.

If one has to look for the dynamics of condensates, it is natural
to use the time-dependent generalization of the Schr\"{o}dinger
equation with the same nonlinear interaction term and obtain the
time-dependent GP equation

\begin{equation}
-\frac{\hbar^2}{2m}\nabla^2 \psi({\bf r})+V({\bf r})\psi({\bf
r})+U_0 |\psi({\bf r})|^2\psi({\bf r})=i\hbar \frac{\partial
\psi({\bf r},t)}{\partial t}.\label{eq1.14}
\end{equation} To ensure consistency between the time-dependent GP equation
(\ref{eq1.14}) and the time-independent GP equation
(\ref{eq1.11}), under stationary conditions $\psi({\bf r},t)$ must
evolve in time as $\exp(-i\mu t/\hbar)$.

In the above equation (\ref{eq1.14}), $\psi({\bf r},t)$, $r =
(x,y,z)$ represents the condensate wave function, $\nabla^2$
denotes the Laplacian operator, $V(\bf{r})$ is the trapping
potential assumed to be $V({\bf r})=m (\omega_r^2 r^2 + \omega_x^2
x^2)$ where $r^2=y^2+z^2$, $\omega_{r,x}$ are the confinement
frequencies in the radial and axial directions respectively, $U_0
= 4\pi \hbar^2 a/m$ corresponds to the strength of interatomic
interaction between the atoms characterized by the short-range
\emph{{s}}-wave scattering length $a$, and  $m$ is the atom mass.

From equation (\ref{eq1.14}) it is obvious that the GP equation is
an inhomogeneous (3+1) dimensional nonlinear Schr\"{o}dinger (NLS)
equation. The inhomogeneity originates from the potential $V({\bf
r})$ that traps the atoms in the ground state and the nonlinearity
coefficient $U_0$ which represents the interatomic interaction
related to the scattering length $a$. The scattering length can
have either positive or negative values which means the
interaction can be either repulsive or attractive. It should be
mentioned that it should be possible to vary the scattering length
periodically with time $a(t)$ employing Feshbach resonance
\cite{inouy98}. This means that understanding the dynamics of BECs
boils down to solving a variable coefficient (3+1) NLS equation
for suitable choices of trapping potentials $V({\bf{r}})$ and
temporal scattering lengths $a(t)$.

\section{Integrability of GP equation- Review of analytical methods}
Looking back at the (3+1) dimensional Gross-Pitaevskii equation
(\ref{eq1.14}), it is obvious that it is in general nonintegrable
for an arbitrary trapping potential and interatomic interaction.
Hence, one has to investigate whether the GP equation would admit
integrability in lower spatial dimensions for specific choices of
trapping potentials and scattering lengths. In other words, one
has to look for the associated nonlinear excitations in (1+1) and
(2+1) dimensional GP equations which would reflect upon the
integrability of the associated dynamical system under
consideration. In a three dimensional BEC, when the transverse
trapping frequency $\omega_r$ ($r=y,z$) is very high compared to
the longitudinal trapping frequency $\omega_x$, then the
transverse confinement is too tight to allow scattering of atoms
to the excited states of the harmonic trap in the transverse
direction. Under this condition, one obtains ``cigar'' shaped BECs
and the three dimensional GP equation becomes quasi
one-dimensional in nature. Again, it should be mentioned that the
quasi one-dimensional GP equation can be shown to be integrable
only for suitable choices of trapping  potential $V(x)$ and
interatomic interaction $U_{0}(x)$.
\begin{center}
\textbf{A.  Analytical Methods}
\end{center}
Eventhough one cannot precisely define the concept of
integrability of a dynamical system governed by a nonlinear
partial differential equation (PDE), one can look for the possible
signatures of integrability namely, Painleve (P-)property
\cite{pproperty}, Lax-pair \cite{lax} soliton solutions etc. A
given nonlinear dynamical system governed by a nonlinear PDE is
said to admit P- property if the corresponding solution can be
locally given in terms of a Laurent series expansion in the
neighborhood of a movable singular point/manifold. The existence
of a Lax-pair of a given nonlinear pde implies that one can
somehow linearize the nonlinear dynamical system and subsequently
exploit it to  generate soliton solutions, thereby consolidating
its integrability. In this section, we dwell upon the analytical
techniques like Inverse Scattering transform \cite{ist}, Gauge
transformation method \cite{llc}, Darboux transformation approach
\cite{darboux,mateev,mateev:2}, Hirota's direct method
\cite{hblm,hblm:2} besides the approximation method like
variational approach.

\begin{center}
\textbf{I.  Inverse Scattering Transform}
\end{center}

\begin{figure}
\begin{center}
\includegraphics[width=0.7\linewidth]{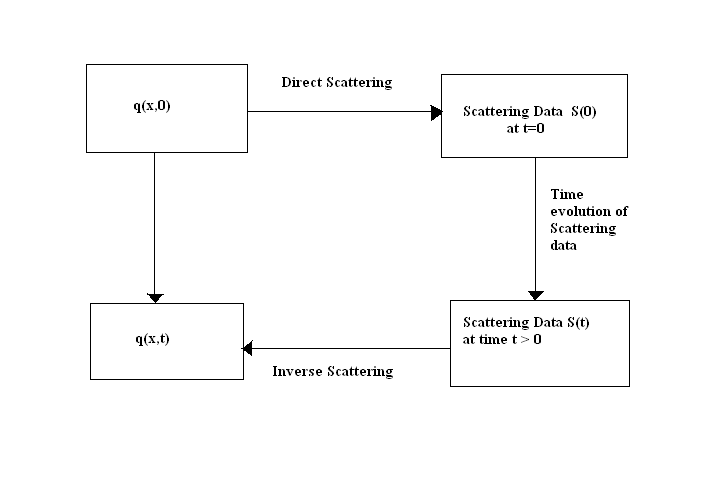}
\caption{Schematic diagram of the inverse scattering transform
method.}\label{ist}
\end{center}
\end{figure}

\quad The Inverse scattering Transform is a nonlinear analogue of
the Fourier transform which has been employed to solve several
linear partial differential equations. Given the initial value of
the potential $q(x,0)$ and the boundary conditions, one has to
identify two linear differential operators $L$ and $B$ so that one
can convert a (1+1) dimensional nonlinear pde into two linear
equations, namely a linear eigenvalue problem
\begin{equation}
L \Phi = \lambda \Phi,\label{eq2.01}
\end{equation}
and a linear time evolution equation
\begin{equation}
\Phi_t  = B \Phi,\label{eq2.02}
\end{equation}
such that the compatability condition of the above two equations
(\ref{eq2.01}) and (\ref{eq2.02}), \emph{{i.e.}}, $L_t = [B,L]$
generates the nonlinear PDE one has started with. Once the
linearization is performed in the above sense for a given
nonlinear dispersive system $q_t = K(q)$, where $K(q)$ is a
nonlinear functional of $q$ and its spatial derivatives, the
Cauchy initial value problem corresponding to the boundary
condition $q\rightarrow 0$ as $|x|\rightarrow \infty$ can be
solved by a three step process indicated schematically in Fig.
\ref{ist}.

This method involves the following three steps: \\  \\
\begin{enumerate}
    \item \textit{Direct scattering transform analysis}

Considering the initial condition $q(x,0)$ as the potential, an
analysis of the linear eigenvalue problem (\ref{eq2.01}) is
carried out to obtain the scattering data $S(0)$. For example, for
the KdV equation we have
\begin{equation}
S(0) = \lambda_n (0), n=1,2, \ldots, N, C_n(0), R(x,0),
-\infty<x<\infty
\end{equation}

where $N$ is the number of bound states with eigenvalues
$\lambda_n$, $C_n(0)$ is the normalization constant of the bound
state eigenfunctions and $R(x,0)$ is the reflection coefficient
for the scattering
data.\\

    \item \textit{Time evolution of scattering data}

Using the asymptotic form of the time evolution equation
(\ref{eq2.02}) for the eigenfunctions, the time evolution of the
scattering data $S(t)$ can be determined.
    \item \textit{Inverse scattering transform (IST) analysis}

The set of $Gelfand-Levitan-Marchenko$ integral equations
corresponding to the scattering data $S(t)$ is constructed and
solved. The resulting solution consists typically of $N$ number of
localized, exponentially decaying solutions asymptotically $(t
\rightarrow \pm \infty)$. In this way, one can successfully solve
the initial value problem of the nonlinear PDE.
\end{enumerate}

From the above, it is obvious that solving the initial value
problem of the given nonlinear PDE boils down to solving an
integral equation.

\begin{center}
\textbf{II.  Gauge Transformation Approach}
\end{center}

This is an iterative method enabling one to generate soliton
solutions starting from a seed solution. In this method, one again
begins with the Zakharov-Shabat(ZS)-Ablowitz-Kaup-Newell-Segur
(AKNS) linear systems \cite{zakha72,ablow73-73} given by the
following equations:
\begin{eqnarray}
\Phi_x = U \Phi \label{evp1}\\
\Phi_t = V \Phi \label{evp2}
\end{eqnarray}
where
\begin{equation}
U = \left(%
\begin{array}{cc}
  -i \lambda & q \\
  r & i \lambda \\
\end{array}%
\right) \hspace{0.5cm}, \hspace{1cm}
V = \left(%
\begin{array}{cc}
  A & B \\
  C & D \\
\end{array}%
\right).\label{uv}
\end{equation}
In the above equation, that is equation (\ref{uv}), $\lambda$
represents the spectral parameter, $(q,r)$ the potential/field
variable of the given nonlinear PDE, $A, B, C$ and $D$ correspond
to the undetermined functions of $\lambda, x, t, q, r$ and their
spatial/time derivatives. The compatibility condition
${(\Phi_{x})}_{t} = {(\Phi_{t})}_{x}$ yields $U_{t} - V_{x} +
[U,V] =$ 0 which is equivalent to the given nonlinear PDE. One
generally calls the matrices $U$ and $V$ as ``Lax-pair'' which
contains information about the linearization of the given
dynamical system. It should be mentioned that for scalar nonlinear
PDEs, $U$ and $V$ are 2$\times$2 matrices and the eigenfunction
$\Phi$ is 2$\times$1 column vector while for vector (two component
in general) nonlinear PDEs, the eigenfunction $\Phi$ is a
3$\times$1 column vector and $U$ and $V$ are 3$\times$3 matrices.

Now, gauge transforming the eigen function $\Phi$, we have an
iterated eigen function $\Phi^{(1)}$=g $\Phi$ where $g$ =
$g(x,t,\lambda)$ is a 2$\times$2 matrix function so that
\begin{eqnarray}
\Phi^{(1)}_x = U^{(1)} \Phi \label{eqphi1}\\
\Phi^{(1)}_t = V^{(1)} \Phi \label{eqphi2}
\end{eqnarray}
where,
\begin{eqnarray}
U^{(1)} = g U g^{-1}+g_x g^{-1},\label{eq2.21} \\
V^{(1)} = g V g^{-1}+g_t g^{-1}.\label{eq2.22}
\end{eqnarray}
The transformation function $g$ must be adopted from the solutions
of certain Riemann problems in the complex $\lambda$ plane and it
must be a meromorphic (regular) function. The simplest form of $g$
satisfying the above criteria can be written as
\begin{equation}
g = \left[I+\frac{\lambda_1-\mu_1}{\lambda-\lambda_1}P(x,t)
\right] \left(%
\begin{array}{cc}
  1 & 0 \\
  0 & -1 \\
\end{array}%
\right),
\end{equation}
where $\lambda_1$ and $\mu_1$ are two arbitrary complex numbers
and $P$ is an undetermined $2\times2$ projection matrix $(P^2 =
P)$. Hence, $g^{-1}$ is now given by
\begin{equation}
g^{-1}= \left(%
\begin{array}{cc}
  1 & 0 \\
  0 & -1 \\
\end{array}%
\right)\cdot \left[I -
\frac{\lambda_1-\mu_1}{\lambda-\mu_1}P(x,t)\right].
\end{equation}

Thus, the vanishing of the apparent residues at $\lambda =
\lambda_1$ and $\lambda =\mu_1$ imposes the following constraints
on $P$ as

\begin{eqnarray}
P_x &=& (I-P) \sigma_3 U (\mu_1) \sigma_3 P - P \sigma_3 U
(\lambda_1) \sigma_3 (I-P),\label{eq2.ph_x}\\
P_t &=& (I-P) \sigma_3 V (\mu_1) \sigma_3 P - P \sigma_3 V
(\lambda_1) \sigma_3 (I-P),\label{eq2.ph_t}
\end{eqnarray}
where
\begin{equation}
\sigma_3 = \left(%
\begin{array}{cc}
  1 & 0 \\
  0 & -1 \\
\end{array}%
\right).
\end{equation}

To generate a new solution from a given solution (vacuum), for
example $q^0$ and $r^0$ associated with matrices $U^0$ and $V^0$,
the eigen value problem takes the following form
\begin{equation}
\Phi_x^0 = U^0 \Phi^0; \quad \quad \Phi_t^0 = V^0
\Phi^0.\label{phinot}
\end{equation}
where $U|_{seed}=U^0$ and $V|_{seed}=U^0$. Now, one can solve
equations (\ref{eq2.ph_x}) and (\ref{eq2.ph_t}) using the vacuum
eigen function $\Phi^0$ such that
\begin{equation}
P= \sigma_3 \tilde{P}\sigma_3,\label{eq2.26}
\end{equation}
where
\begin{equation}
\tilde{P}=M^{(1)}/[\rm trace \it M^{(\rm 1)}].
\end{equation}
and  $M^{(1)}$ is a 2$\times$2 matrix  defined by
\begin{equation}
M^{(1)} \equiv \Phi^0(x,t;\mu_1)\left(%
\begin{array}{cc}
  m_1 & 1/n_1 \\
  n_1 & 1/m_1 \\
\end{array}%
\right)\Phi^0(x,t;\lambda_1)^{-1},
\end{equation}
where $m_1$ and $n_1$ are arbitrary complex constants and
$\Phi^{0}$ is a solution of the vacuum linear system governed by
equations (\ref{phinot}).

Now, substituting the meromorphic function $g$ with the projection
matrix $P$ given by equation (\ref{eq2.26}) into equation
(\ref{eq2.21}), we get
\begin{equation}
U'(\lambda) = \left(%
\begin{array}{cc}
  -i \lambda  & -q^0  \\
  -r^0 & i\lambda \\
\end{array}%
\right)-2i(\lambda_1-\mu_1)\left(%
\begin{array}{cc}
  0 & \tilde{P}_{12} \\
  -\tilde{P}_{21} & 0 \\
\end{array}%
\right).
\end{equation}

A comparison of the eigenvalue problems expressed in terms of the
vacuum eigenfunction $\Phi^{0}$ and the new (transformed)
eigenfunction $\Phi^{(1)}$ would enable us to relate the vacuum
solution of the associated nonlinear PDE with the new solution.
Thus, we get an explicit solution as \cite{llc}
\begin{eqnarray}
q^{(1)} =  -q^0 -2i (\lambda_1 -\mu_1) \tilde{P}_{12},\label{eq2.30}\\
r^{(1)} =  -r^0 +2i (\lambda_1 -\mu_1)
\tilde{P}_{21},\label{eq2.31}
\end{eqnarray}
where
\begin{equation}
\tilde{P}_{12}=\frac{M_{12}^{(1)}}{M_{11}^{(1)}+M_{22}^{(1)}},\quad
\quad
\tilde{P}_{21}=\frac{M_{21}^{(1)}}{M_{11}^{(1)}+M_{22}^{(1)}}.
\end{equation}

Once we get $q^{(1)}$ and $r^{(1)}$ from the input solution $q^0$
and $r^0$, one can repeat the same procedure to obtain yet another
new solution $q^{(2)}$ and $r^{(2)}$) using $q^{(1)}$ and
$r^{(1)}$) as the input solution. For example, to construct second
iterated solution $q^{(2)}$, $r^{(2)}$, one needs to find a
solution of the linear system given by equations (\ref{eqphi1})
and (\ref{eqphi2}) where the matrices $U^1$ and $V^1$ are
associated with the input solutions $q^{(1)}$ and $r^{(1)}$.
However, one can find the solution of above eigen value problem in
terms of $\Phi^{0}$ as
\begin{equation}
\Phi^{(1)}=g \Phi^0.
\end{equation}

Thus, the new iterated solution $q^{(2)}$ and $r^{(2)}$ (in
analogy with equations (\ref{eq2.30}) and (\ref{eq2.31})) can be
written as
\begin{eqnarray}
q^{(2)} = -q^{(1)}-2i(\lambda_2 - \mu_2)\tilde{P}_{12}^{(1)},\\
r^{(2)} = -r^{(1)}+2i(\lambda_2 - \mu_2)\tilde{P}_{21}^{(1)}.
\end{eqnarray}

Thus, one can repeat the same procedure $N$-times to obtain the
$N^{\rm{th}}$ iterated solution as

\begin{eqnarray}
q^{N} = - q^{(N-1)}-2i (\lambda_N - \mu_N)\tilde{P}^{(N-1)}_{12},\\
r^{N} = - r^{(N-1)}-2i (\lambda_N - \mu_N)\tilde{P}^{(N-1)}_{21}.
\end{eqnarray}

The above iteration could become extremely handy, particularly in
the context of the generation of multisoliton solutions as one can
obtain $N$-soliton solution from the vacuum eigen function
$\Phi^{0}$ of the linear system.
\begin{center}
\textbf{III.  Darboux Transformation Approach}
\end{center}
In 1882 G. Darboux  studied the eigen value problem of the one
dimensional Schr\"{o}dinger equation

\begin{equation}
-\Phi_{xx}-q(x) \Phi  = \lambda \Phi,\label{eq2.1}
\end{equation}
where $q(x)$ is a potential function and $\lambda$ is a constant
spectral parameter. He postulated that if $q(x)$ and
$\Phi(x,\lambda)$ are two functions satisfying equation
(\ref{eq2.1}) and $f(x) = \Phi(x, \lambda_0)$ is a solution of
equation of (\ref{eq2.1}) for $\lambda = \lambda_0$ where
$\lambda_0$ is a fixed constant, the functions $q'$ and $\Phi'$
are defined by
\begin{equation}
q'=q+2(\ln f)_{xx}, \quad\quad \Phi'(x,\lambda)=\Phi_x (x,\lambda)
- \frac{f_x}{f}\Phi(x,\lambda),\label{eq2.2}
\end{equation}
with
\begin{equation}
-\Phi'_{xx}-q'\Phi' = \lambda \Phi'.\label{eq2.3}
\end{equation}

From equations (\ref{eq2.1}) and (\ref{eq2.3}), it is obvious that
they are of the same form. Therefore, the transformation
(\ref{eq2.2}) which converts the functions ($q, \Phi$) to ($q',
\Phi'$) satisfying the same partial differential equation is the
original Darboux transformation which is valid for $f\neq 0$.

In this method, one again begins with the ZS-AKNS linear eigen
value problem given by equations (\ref{evp1}) and (\ref{evp2}),
where the eigen function $\Phi$ is a 2$\times$2 matrix of the
following form
\begin{equation}
\Phi(x,t,\lambda) = \left(%
\begin{array}{cc}
  \Phi_{11}(x,t,\lambda) & \Phi_{12}(x,t,\lambda) \\
  \Phi_{21}(x,t,\lambda) & \Phi_{22}(x,t,\lambda) \\
\end{array}%
\right).
\end{equation}

Now, introducing the Darboux transformation into the known eigen
function, we now obtain the transformed eigenfunction as
\begin{equation}
\Phi^{(1)}(x,t,\lambda) = D(x,t,\lambda)
\Phi(x,t,\lambda),\label{eq2.9}
\end{equation}
where $D(x,t,\lambda)$ is the ``Darboux matrix'' which is
equivalent to $\lambda I - S$. While $I$ is the identity matrix,
the matrix $S$ can be generated as
\begin{equation}
S = H \Lambda H^{-1}, \quad \rm{det} H \neq 0,
\end{equation}
where the matrix $H$ is defined as $H = (h_1,\ldots,h_N)$, where
$h_i$ represents the column solution of the linear eigenvalue
problem given by equations (\ref{evp1}) and (\ref{evp2}), $\Lambda
=\rm{diag}(\lambda_1,\ldots,\lambda_N)$. The new eigen function
$\Phi^{(1)}$ again satisfies equations (\ref{eqphi1}) and
(\ref{eqphi2}) and the Darboux matrix $D$ plays the role of the
transformation function $g$. Accordingly, we have

\begin{eqnarray}
U^{(1)} = D U D^{-1} + D_x D^{-1},\label{eq2.12}\\
V^{(1)} = D U D^{-1} + D_t D^{-1}.\label{eq2.13}
\end{eqnarray}

To check the form of $U^{(1)}$ given by equation (\ref{eq2.12}),
we substitute eqns. (\ref{evp1}), (\ref{eqphi1}) into the $x$
derivative of equation(\ref{eq2.9}) to obtain
\begin{equation}
U^{(1)} \Phi^{(1)} = D_x \Phi +D U \Phi.\label{eq2.14}
\end{equation}

Then, making use of equation (\ref{eq2.9}), one obtains equation
(\ref{eq2.12}). Similarly, one can check the form of $V^{(1)}$
given by (\ref{eq2.13}).

Thus, starting from a seed solution $U$ of the given nonlinear
pde, one can generate the iterated $U^{(1)}$. This procedure can
be repeated to generate multisoliton solutions.
\begin{center}
\textbf{IV. Hirota Bilinear Method}
\end{center}

Although the inverse scattering formalism was the first analytical
technique that has been developed to solve the initial value
problem of nonlinear pdes, it involves sophisticated mathematical
concepts like solving an integral equation and hence quite
complicated and intricate. Moreover in this method, one should
have a prior knowledge of the potential $u(x,t)$ at $t=0$, namely
the initial data $u(x,0)$ and the boundary conditions imposed on
it. On the other hand, eventhough Darboux and gauge transformation
approaches are iterative in nature and are purely algebraic
without involving highly complex mathematics, they warrant the
identification of the Lax-pair of the associated dynamical system.
Hence, it is imperative to look for an alternative method to
generate localized solutions (soliton solutions) of nonlinear pdes
and in this context, Hirota's direct method comes to our rescue.
In this method, neither does one need any prior information about
the potential (or physical field) of the associated nonlinear pde,
nor the Lax-pair of the associated dynamical system. This method
which has an inbuilt deep algebraic and geometric structure is
more elegant and straightforward and can be directly employed to
generate soliton solutions of nonlinear pdes.

The salient features of the Hirota method are the following:\\
i) The given nonlinear partial differential equation has to be
converted into a bilinear equation through a transformation which
can be identified from the Painlev\'e analysis. Each term
of the bilinear equation has the degree two.\\

The Hirota bilinear operators are defined as

\begin{eqnarray}
D_t^m D_x^n(G\cdot F)=\left(\frac{\partial}{\partial
t}-\frac{\partial}{\partial t'}\right)^m
\left(\frac{\partial}{\partial x}-\frac{\partial}{\partial
x'}\right)^n \notag \\G(t,x)F(t',x')\notag |_{t'=t, x'=x}.\notag
\end{eqnarray} ii) The dependent variables $G$ and $F$ in the bilinear form have
to be expanded in the form of a power series in terms of a small
parameter $\varepsilon$ as
\begin{eqnarray}
G&=&\varepsilon g^{(1)} +\varepsilon^3 g^{(3)} +\varepsilon^5
g^{(5)}+\cdots ,\\
F&=&1+\varepsilon^2 f^{(2)} +\varepsilon^4 f^{(4)}+\cdots.
\end{eqnarray} iii) After substituting the above functions into the bilinear form
and equating different powers of $\varepsilon$, a set of linear
pdes can be generated. \\ iv) Finally, solving the linear pdes,
one can generate soliton solutions.

It should be mentioned that the key to the success of the method
lies in the identification of the dependent variable
transformation as well as in choosing an optimum power series to
linearise the given nonlinear pde.

\subsection{Approximation method}
\begin{center}
\textbf{I.  Variational Approach}
\end{center}
Variational approach is a qualitative semi-analytical approach. By
using the variational approximation, one can study the dynamics of
Bose-Einstein condensates described by the mean-field
Gross-Pitaevskii equation. For the purpouse of variational
analysis, Larangian density is calculated for the corresponding
time-dependent Gross-Pitaevskii  equation and the effective
Lagrangian can be obtained by integrating the initial trial wave
function with variational parameters over space. The numerical
value of each variational parameter can be obtained from the
numerical solution of corresponding Euler-Lagrangian equations.

\section {Bright matter wave solitons and their collision in Scalar BECs
in certain simple potentials}

In this section, we investigate the dynamics of scalar
Bose-Einstein condensates in certain simple physically realizable
potentials. To start with, we consider the dynamics of BECs in a
time independent harmonic trap with exponentially varying
scattering length. We generate the associated bright solitons and
study their collisional dynamics. We then introduce suitable time
dependence in the harmonic trap and investigate the dynamics of
BECs. We then show that how the interplay between trap frequency
and temporal scattering length can generate matter wave
interference pattern in the collision of bright solitons. We then
reinforce the binary attraction with a repulsive three body
interaction to enhance the stability of BECs.

\begin{center}
\textbf{A.  Dynamics of quasi one dimensional BECs in a time
independent harmonic trap}
\end{center}

For a time independent harmonic oscillator potential and
exponentially varying scattering lengths, the GP equation
(\ref{eq1.14}) for cigar shaped BECs takes the following form
\cite{perez98,salas02-65,liang05,khawa06,radha07}
\begin{equation}
i\frac{{\partial \psi }}{{\partial t}} + \frac{{\partial ^2 \psi
}}{{\partial x^2 }} + 2\;a(t)\;\left| {\psi } \right|^2 \psi
+\frac{1}{4}\lambda ^2 x^2 \psi  = 0,\label{ch4eq2}
\end{equation}
where time $t$ and coordinate $x$ are measured in units
$2/\omega_\bot$ and $a_\bot$, where $a_\bot = (\hbar/m
\omega_\bot)^{1/2}$ and $a_0 = (\hbar/m\omega_0)^{1/2}$ are linear
oscillator lengths in the transverse and cigar-axis direction
respectively. $\omega_\bot$ and $\omega_0$ are corresponding
harmonic oscillator frequencies, $m$ is the atomic mass and the
trap frequency $\lambda = 2|\omega_0|/\omega_\bot << 1$. The
Feshbach resonance managed nonlinear coefficient which represents
the scattering length reads $a(t) = \tilde{a}_0$exp$(\lambda t)$.

Equation (\ref{ch4eq2}) admits the following Lax-pair
\cite{khawa06}
\begin{subequations}
\begin{eqnarray}
\Phi_{x}&=&U \Phi,\;\;\;\;U= \left({\begin{array}{*{20}c}
   {i\zeta} & Q  \\
   {-Q{*}} & {-i \zeta} \\
\end{array}} \right), \label{ch4eq3a}\\
\Phi_{t}&=&V\Phi, \label{ch4eq3b}\\
V&=& \left({\begin{array}{*{20}c}
   {-2i\zeta^2+i\lambda x \zeta+i \left| Q \right|^2} &  [(\lambda x -2 \zeta)Q+iQ_{x}]  \\
   { [-(\lambda x -2 \zeta)Q^{*}+iQ^{*}_{x}]} & {{2i\zeta^2-i\lambda x \zeta-i  \left| Q \right|^2}} \\
\end{array}} \right),\nonumber
\end{eqnarray}
\end{subequations}
where we have slightly modified the Lax-pair (given in ref.
\cite{khawa06}) by allowing the nonisospectral parameter $\zeta$
to be complex keeping the initial scattering length unity. The
nonisospectral complex parameter obeys the first order ordinary
differential equation of the form
\begin{equation}
\zeta_{t}=\lambda
\zeta,\;\;\;\;\zeta(t)=\alpha(t)+i\beta(t),\label{ch4eq4}
\end{equation}
and the macroscopic wave function $\psi$ is related to $Q$ by the
transformation
\begin{equation}
Q=exp\left( \frac{{\lambda t}}{2} + i\frac{{\lambda x^2
}}{4}\right )\psi (x,t).\label{ch4eq5}
\end{equation}

Now, to generate the bright soliton solutions of equation
(\ref{ch4eq2}), we consider a trivial vacuum solution $Q^{(0)}=0$
to give the following vacuum linear systems
\begin{subequations}
\begin{eqnarray}
\Phi_{x}^{(0)}&=&\left({\begin{array}{*{20}c}
   {i\zeta}& 0 \\
   {0}&{-i \zeta}\\
\end{array}}\right)\Phi^{(0)}=U^{(0)}\Phi^{(0)}, \label{ch4eq6a}\\
\Phi_{t}^{(0)} & = & \left({\begin{array}{*{20}c}
   {-2i\zeta^2+i\lambda x \zeta} & 0  \\
   {0} & {{2i\zeta^2-i\lambda x \zeta}} \end{array}} \right)\Phi^{(0)}\notag \\
   & = & V^{(0)} \Phi^{(0)}.\label{ch4eq6b}
\end{eqnarray}
\end{subequations}

Solving the above linear systems keeping in mind that the spectral
parameter $\zeta$ varies with time by virtue of equation
(\ref{ch4eq4}), we have
\begin{equation}
\Phi^{(0)}(x,t,\zeta)= \left({\begin{array}{*{20}c}
   {e^{i x \zeta-2i\int\zeta^2 dt}} & 0  \\
   {0} & {{e^{-i x \zeta+2i\int\zeta^2 dt}}} \\
\end{array}} \right).\label{ch4eq7}
\end{equation}

Now, effecting the gauge transformation
\begin{equation}
\Phi^{(1)}(x,t,\zeta)=g  \Phi^{(0)}(x,t,\zeta),\label{ch4eq8}
\end{equation}
where ``$g$'' is a meromorphic solution of the associated Riemann
problem. The new linear eigenvalue problems now take the following
form
\begin{equation}
\Phi_x^{(1)}=U^{(1)} \Phi^{(1)},\;\;\;\;\Phi_t^{(1)}=V^{(1)}
\Phi^{(1)},\label{ch4eq9}
\end{equation}
with
\begin{subequations}
\begin{eqnarray}
U^{(1)}=g U^{(0)} g^{-1} + g_x g^{-1},\label{ch4eq10a}\\
V^{(1)}=g V^{(0)} g^{-1} + g_t g^{-1}.\label{ch4eq10b}
\end{eqnarray}
\end{subequations}
We now choose $g$ as
\begin{equation} g= \left(1 +
\frac{{\zeta_{1}-\mu_{1}}}{\zeta -
\zeta_1}P_{1}(x,t)\right)\left({\begin{array}{*{20}c}
   {1} & 0  \\
   {0} & {-1} \\
\end{array}} \right)\label{ch4eq11}
\end{equation} where $\zeta_1$ and $\mu_1$ are arbitrary complex parameters and
$P_1$ is a projection matrix $(P_1^2=P_1)$. Imposing the
constraint that $U^{(1)}$ and $V^{(1)}$ do not develop
singularities  around the poles $\zeta = \zeta_1$ and
$\zeta=\mu_1$, the choice of the projection matrix $P_1$ is
governed by the solution of the following set of partial
differential equations
\begin{subequations}
\begin{eqnarray}
P_{1x}&=&(1-P_{1})\sigma_{3}U^{(0)}(\mu_{1})\sigma_{3}P_{1}-P_{1}\sigma_{3}U^{(0)}(\zeta_{1})\notag\\&&\sigma_{3}(1-P_{1})\label{ch4eq12a}\\
P_{1t}&=&(1-P_{1})\sigma_{3}V^{(0)}(\mu_{1})\sigma_{3}P_{1}-P_{1}\sigma_{3}V^{(0)}(\zeta_{1})\notag\\&&\sigma_{3}(1-P_{1})\label{ch4eq12b}
\end{eqnarray}
\end{subequations} where

\begin{equation}
\sigma_{3}=\left({\begin{array}{*{20}c}
   1 & 0  \\
   0 & -1 \\
\end{array}} \right).
\end{equation}

Looking at the above system of equations, we understand that $P_1$
depends only on the trivial matrix eigenfunction $\Phi^{(0)}
(x,t,\zeta)$, a
diagonal matrix and has a compact form given by\\
\begin{subequations}
\begin{eqnarray}
P_1 (x,t)&=&\sigma_{3} \frac{{M^{(1)}}}{[\rm trace\;M^{(1)}]}\sigma_3,\label{ch4eq13a}\\
M^{(1)}&=& \left({\begin{array}{*{20}c}
   m_1 & 1/n_1  \\
   n_1 & {1/m_1}  \\
\end{array}} \right)\Phi^{(0)}(x,t,\zeta_1)^{-1}\label{ch4eq13b}
\end{eqnarray}
\end{subequations}
where $m_1$ and $n_1$ are arbitrary complex constants. Hence,
choosing the complex parameters $\zeta_1=\alpha_1 (t)+i\beta_1
(t)$  and $\mu_1=\zeta_{1}^*$ and employing the gauge
transformation approach \cite{llc}, we arrive at the matter wave
bright soliton
\begin{equation}
\psi^{(1)} (x,t)=2 \beta_{0} exp\left( \frac{{\lambda t}}{2} -
\frac{{i\lambda x^2 }}{4}\right
)sech(\theta_1)exp(i\xi_{1}),\label{ch4eq14}
\end{equation}
where
\begin{subequations}
\begin{eqnarray}
\theta_1&=&2\beta_{1}x -8\int (\alpha_{1}\beta_{1})dt+2\delta_1,\label{ch4eq15a}\\
\xi_1&=&2\alpha_{1}x -4\int(\alpha_{1}^2 - \beta_{1}^2)dt-2\phi_1,\label{ch4eq15b}\\
\alpha_1&=&\alpha_{1 0}e^{\lambda t},\;\;\;\;\beta_1=\beta_{1
0}e^{\lambda t}.\label{ch4eq15c}
\end{eqnarray}
\end{subequations}

\begin{figure}
\begin{center}
\includegraphics[width=0.4\linewidth]{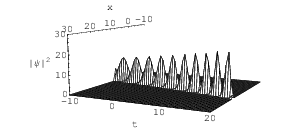}

\caption{The dynamics of bright solitons for the parametric choice
$\lambda$= 0.02, $\beta_{0}$=2, $\alpha_0$=0.1, $\delta_1$=0.5,
$\phi_1$=0.1.}\label{in.1}
\end{center}
\end{figure}

\begin{figure}
\begin{center}
\includegraphics[width=0.4\linewidth]{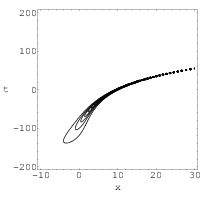}
\caption{Contour Plot of bright soliton for
$\lambda$=0.02.}\label{in.2}
\end{center}
\end{figure}

\begin{figure}
\begin{center}
\includegraphics[width=0.4\linewidth]{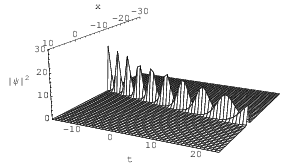}
\caption{The dynamics of bright soliton for $\lambda$=
-0.02.}\label{in.3}
\end{center}
\end{figure}

\begin{figure}
\begin{center}
\includegraphics[width=0.4\linewidth]{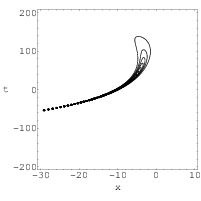}
\caption{Contour plot of bright soliton for $\lambda$=
-0.02.}\label{in.4}
\end{center}
\end{figure}

The above bright soliton solution given by equations
(\ref{ch4eq14}) and (\ref{ch4eq15a}-\ref{ch4eq15c}) is identical
to the one given by Liang \emph{{et al.}} \cite{liang05} using
Darboux transformation. From the  profile of the bright soliton
trains shown in Fig. \ref{in.1}, we infer that matter wave density
$|\psi|^2$ increases with the increase of the absolute value of
the scattering length leading to the compression of the bright
soliton trains of BEC. The contour plot (shown in Fig. \ref{in.2})
which takes a cross-sectional view of Fig. \ref{in.1} in the $x-t$
plane shows that the width of bright solitons decreases
progressively. Figure \ref{in.3} shows that the peak value of the
matter wave density $|\psi| ^2$ decreases with the decrease of the
absolute value of the scattering length leading to a broadening of
the bright soliton trains thereby enhancing their width and this
is again confirmed by the corresponding contour plot in figure
\ref{in.4}. Our investigation shows that the scattering length can
be suitably manipulated to compress the bright solitons of BECs
into an assumed peak matter density without causing their
explosion while on the other hand, it can be manoeuvred
judiciously to broaden the localized solitons without allowing the
dilution of the condensates and this interpretation completely
agrees with that of Liang \emph{{et al.}} \cite{liang05}.
Investigations of the quasi one dimensional GP equation
\cite{khawa06,liang05,chong07,li06,kengn06,yuce07,wu07} in the
presence of an expulsive parabolic potential for positive
scattering lengths also confirm our above observations.

From the above, we observe that one can either compress or broaden
the bright solitons in the expulsive time independent trap either
by exponentially increasing or decreasing the scattering lengths
respectively. It must be mentioned that the exponentially varying
scattering length with a trap frequency dependence would make the
exact solutions of the GP equation less interesting from an
experimental point of view. Hence, it would be interesting to
investigate the impact of a general time dependent scattering
length and a time dependent trap on the condensates. The addition
of time dependent trap frequency will facilitate us to tune the
trap suitably and study its impact on the condensates.

\begin{center}
\textbf{B.  Impact of transient trap on  BECs}
\end{center}

The introduction of time dependance in the trap ensures that the
condensates are now confronted with both time dependent scattering
length and time dependent trapping potential and accordingly
equation (\ref{ch4eq2}) gets modified as (in dimensionless units)
\cite{atre06}

\begin{equation}
i\frac{\partial \psi}{\partial
t}+\frac{1}{2}\frac{\partial^2\psi}{\partial x^2}+a(t) |\psi|^2
\psi -\frac{\lambda(t)}{2}x^2 \psi = 0,\label{ch4-3eq2}
\end{equation}
where $a(t) = -2 a_s (t) / a_B$, $\lambda(t)=\omega_0^2 (t)/
\omega^2_\perp$, $a_B$ is the Bohr radius, $\lambda(t)$ describes
time dependent harmonic trap which can be confining ($\lambda(t) <
0$) or expulsive ($\lambda(t) > 0 $). This equation has been
mapped onto a linear Schr$\rm \ddot{o}$dinger eigenvalue problem
\cite{atre06} and one soliton solution has been expressed in terms
of doubly periodic Jacobian elliptic functions.  However, the
above approach cannot be employed to generate multisoliton
solutions analytically. The gauge transformation approach comes in
handy at this juncture as it offers the advantage of constructing
multisoliton solution from the solution of the corresponding
vacuum linear system.

Now, to generate the bright solitons of equation (\ref{ch4-3eq2})
for both regular and expulsive potentials, we introduce the
following modified lens transformation
\cite{atre06,theoc03-67,sulem99,wu07}
\begin{equation}
\psi(x,t)=\sqrt{A(t)} Q(x,t) \rm exp(i \Phi(x,t)),\label{ch4-3eq3}
\end{equation}
where the phase has the following simple quadratic form
\begin{equation}
\Phi(x,t) = - \frac{1}{2}c(t) x^2.\label{ch4-3eq4}
\end{equation}

Substituting the modified lens transformation given by equation
(\ref{ch4-3eq3}) in equation (\ref{ch4-3eq2}), we obtain the
modified NLS equation
\begin{equation}
iQ_{t}+\frac{1}{2}Q_{xx}-ic(t)xQ_{x}-ic(t)Q+a(t)A(t)|Q|^{2}Q=0,\label{ch4-3eq5}
\end{equation}
with
\begin{equation}
\lambda(t) = c'(t) - c(t)^2,\label{ch4-3eq6}
\end{equation}
and
\begin{equation}
c(t) = -\frac{d}{dt}\rm ln A(t).\label{ch4-3eq7}
\end{equation}

Equation (\ref{ch4-3eq5}) admits the following linear eigenvalue
problem
\begin{eqnarray}
\phi_x &=& U \phi,\qquad U=\left(
                           \begin{array}{cc}
                             i \zeta(t) & Q \\
                             -Q^* & -i \zeta (t) \\
                           \end{array}
                         \right),\label{ch4-3eq8}\\
\phi_t &=& V \phi,\label{ch4-3eq9}\\
                     V &=& \left(\begin{array}{cc}
                                  V_{11} & V_{12}\\\\
                                  V_{21} &  V_{22}\\
                                \end{array}
                              \right).\nonumber
\end{eqnarray}
where
\begin{eqnarray}
V_{11}&=& - i \zeta(t)^2+ i c(t) x \zeta(t)+ \frac{i}{2}a(t)A(t)|Q|^2 \nonumber\\
V_{12}&=& (c(t)x-\zeta(t))Q +\frac{i}{2}Q_x \nonumber\\
V_{21}&=& -(c(t)x-\zeta(t))Q^* +\frac{i}{2}Q^*_x \nonumber\\
V_{22}&=& i \zeta(t)^2-i c(t) x \zeta(t)- \frac{i}{2}a(t)A(t)|Q|^2
\end{eqnarray}

 In the above linear eigenvalue problem, the
spectral parameter ``$\zeta$'' which is complex is nonisospectral
obeying the following equation
\begin{equation}
\zeta'(t) = c(t) \zeta (t),\label{ch4-3eq10}
\end{equation}
with $a(t) = 1/A(t)$. It is obvious that the compatability
condition  $(\phi_x)_t = (\phi_t )_x$ generates equation
(\ref{ch4-3eq5}).

Substituting equation (\ref{ch4-3eq7}) with $a(t) = 1/A(t)$ in
equation (\ref{ch4-3eq6}), we get
\begin{equation}
a''(t)a(t)-2a'(t)^2-\lambda(t)a(t)^2=0.\label{ch4-3eq11}
\end{equation}

From the above, it is evident that the GP equation
(\ref{ch4-3eq2}) is completely integrable only if the trap
frequency $\lambda(t)$ and the scattering length $a(t)$ are
connected by equation (\ref{ch4-3eq11}) and the above condition is
consistent with Ref. \cite{wu07}. Thus, the modified lens
transformation has facilitated the identification of integrability
of equation (\ref{ch4-3eq2}). It should also be mentioned that
equation (\ref{ch4-3eq2}) is completely integrable only for
certain suitable choices of trap frequency $\lambda(t)$ depending
on the solvability of equation (\ref{ch4-3eq6}). For example, when
$\lambda(t)={\rm{constant}}=k$ and $a(t)$=$e^{\Lambda t}$, where
$\Lambda$ is the trap frequency, equation (\ref{ch4-3eq2}) reduces
to (\ref{ch4eq2}) describing the dynamics of BECs moving in an
expulsive parabolic potential and exponentially varying scattering
length \cite{liang05,radha07}. The above parametric choice is
consistent with equation (\ref{ch4-3eq11}) with $k=-\lambda^2$
ensuring the integrability of the model. The above model has also
been experimentally realized \cite{khayko}.

To generate the soliton solution of equation (\ref{ch4-3eq5}) (or
equation (\ref{ch4-3eq2})), we consider the seed solution
$Q^{(0)}=0$ and solve the linear systems given by equations
(\ref{ch4-3eq8}) and (\ref{ch4-3eq9}) keeping in mind equation
(\ref{ch4-3eq10}) to obtain

\begin{equation}
\phi^{(0)}(x,t,\zeta)=\left(
  \begin{array}{cc}
    e^{i x \zeta(t)-i \int^t_0 \zeta(t)^2 dt} & 0 \\
    0 & e^{-i x \zeta(t)+i\int^t_0 \zeta(t)^2 dt} \\
  \end{array}
\right).\label{ch4-3eq14}
\end{equation}

Employing the gauge transformation approach and choosing
$\zeta_{1}=\alpha_{1}(t)+i\beta_{1}(t)$ and
$\mu_{1}=\zeta_{1}^{*}$, one obtains the one soliton solution of
equation (\ref{ch4-3eq2}) \cite{rames08pra}
\begin{equation}
\psi^{1}(x,t) = \sqrt{\frac{1}{a (t)}} 2 \beta_1 (t)\;\;
\rm{sech}\theta_1\;\; e^{-\frac{i}{2} c(t) x^2 + i \xi_1
},\label{ch4-3eq23}
\end{equation}
where
\begin{eqnarray}
\theta_1 &=& 2 \beta_1 (t) x - 4\int^t_0{(\alpha_1 (t^{'})\beta_1 (t^{'}))}dt^{'} + 2\delta_1, \nonumber \\
\xi_1 &=& 2 \alpha_1 (t) x - 2\int^t_0{(\alpha_1 (t^{'})^2-\beta_1
(t^{'})^2)}dt^{'} - 2 \phi_1, \\\label{ch4-3eq22}
\alpha_{1}&=& \alpha_{10} e^{\int^t_0 c(t^{'}) dt^{'}},\nonumber\\
\beta_{1}&=&\beta_{10} e^{\int^t_0 c(t^{'}) dt^{'}},\nonumber
\end{eqnarray}
and $\phi_1$, $\delta_1$, $\alpha_{10}$ and $\beta_{10}$ are
arbitrary real constants.

The striking feature of this bright soliton solution is that its
amplitude relies strongly on the scattering length $a(t)$ and the
time dependent trap $\lambda(t)$ while the velocity is governed by
the external trap $\lambda(t)$ alone.

This procedure can be easily extended to generate multisoliton
solution and one can study the collisional dynamics of bright
solitons for a suitable choice of $\lambda(t)$ and $a(t)$
consistent with equation (\ref{ch4-3eq11}).

Thus, it is obvious that one can obtain varieties of soliton
profiles depending on the choice of the scattering length $a(t)$
and the time dependent trap $\lambda(t)$ consistent with equation
(\ref{ch4-3eq11}).

\begin{figure}
\begin{center}
\includegraphics[width=0.65\linewidth]{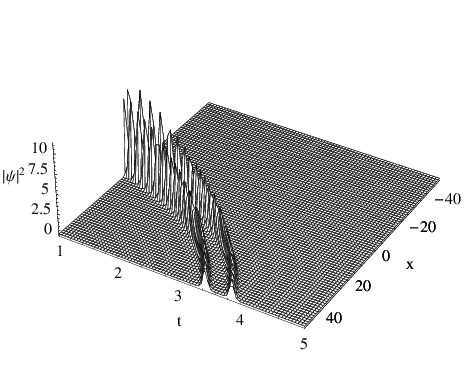}

\caption{Two soliton interaction in the expulsive trap
($\lambda(t)<0$) with $a(t)$ =$a_{0} exp(-0.125 t^2)$,
$a_{0}=0.5$, $\alpha_{10}=2.31$, $\beta_{10}=1.5$,
$\alpha_{20}=3.12$, $\beta_{20}=1.2$,
 $\phi_{1}=.005$, $\delta_{1}=0.002$,
$\phi_{2}=0.002$, $\delta_{2}=0.001$.}\label{td.1}
\end{center}
\end{figure}

\begin{figure}
\begin{center}
\includegraphics[width=0.6\linewidth]{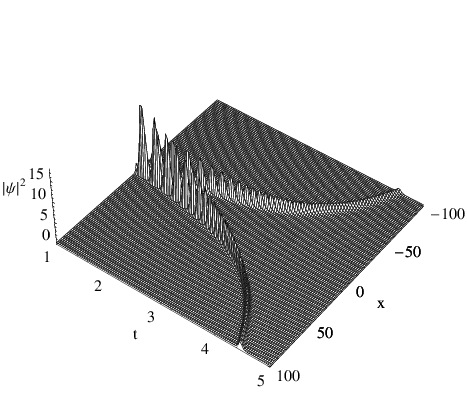}
\caption{Two soliton interaction in the expulsive trap
($\lambda(t)<0$) with $a(t)$ =$a_{0} exp(-0.125 t^2)$,
$a_{0}=0.5$, $\alpha_{10}=2.31$, $\beta_{10}=1.5$,
$\alpha_{20}=-2.12$, $\beta_{20}=1.2$, $\phi_{1}= .05$,
$\delta_{1}=0.02$, $\phi_{2}=0.02$, $
\delta_{2}=0.01$.}\label{td.2}
\end{center}
\end{figure}

Figures (\ref{td.1}) and (\ref{td.2}) describe the evolution of
the two soliton solution for an expulsive trap ($\lambda(t)<0$)
for different initial conditions evolving the scattering length of
the form $a(t)=0.5\exp(-0.125 t^2)$. From the figures, one
observes that the matter wave density $|\psi|^2$ of the
condensates decreases slowly by virtue of the decrease in the
absolute value of the scattering length and the trajectory of the
soliton pulses is dictated by the initial conditions. It should be
mentioned that the identification of this critical parametric
regime in which one observes the slow decay of the condensates
enables one to avoid this domain by operating the system under a
safe range of parameters.

\begin{figure}
\includegraphics[width=0.6\linewidth]{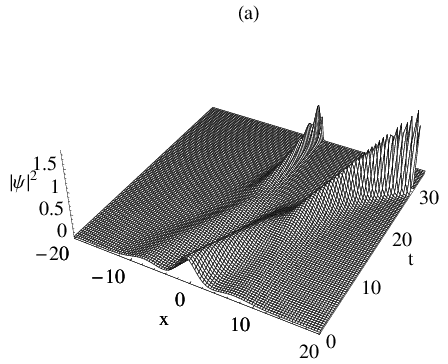}
\includegraphics[width=0.4\linewidth]{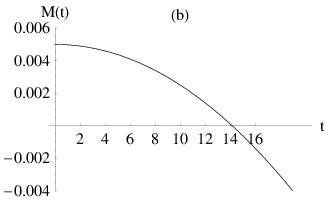}
\includegraphics[width=0.5\linewidth]{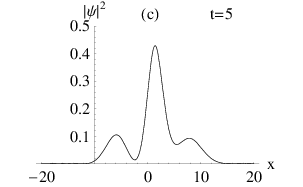}
\includegraphics[width=0.5\linewidth]{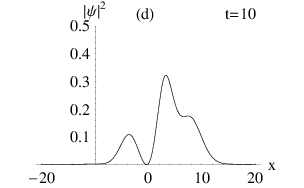}

\caption{Two soliton interaction in the confining trap
($\lambda(t)>0$) at different intervals of time with $a(t)$
=$a_{0}$ $\exp(0.0025 t^2)$, $a_{0}=0.5$, $\alpha_{10}=0.01$,
$\beta_{10}=0.1$, $\alpha_{20}=0.28$, $\beta_{20}=0.11$,
$\phi_{1}= \delta_{2}=0.1$, $\phi_{2}=
\delta_{1}=0.2$.}\label{td.3}
\end{figure}

Figure \ref{td.3} (a) shows the interaction of solitons for $a(t)
=0.5\exp(0.0025 t^2)$. It can be observed from fig. (\ref{td.3}b)
that the confining nature of the trap ($\lambda(t)>0$) is
preserved only for a finite length of time ($t<14$). During this
period, the two soliton pulses slide over each other like liquid
balls as shown in figures \ref{td.3} (c) and \ref{td.3} (d). After
this critical period ($t>$14), the trap becomes expulsive again
which sets in the compression of the soliton pulses resulting in
the increase of the matter wave density $|\psi|^2$ of the
condensates. It can also be observed that for $a(t)
=0.5\exp(-0.125 t^2)$, the absolute value of the scattering length
decreases and hence there is a slow decay of the condensates while
for $a(t) =0.5\exp(0.0025 t^2)$, eventhough the absolute value of
the scattering length increases, the soliton pulses begin to get
compressed (or the matter wave density $|\psi|^2$ increases) after
a finite time delay. It can be easily understood that this delay
is introduced by the time dependent trap. When the time dependent
trap $\lambda(t)$ becomes a constant, equation (\ref{ch4-3eq2})
reduces to the dynamics of BECs in an expulsive parabolic
potential and time independent scattering length. Under this
condition, the soliton trains begin getting compressed and the
matter wave density $|\psi|^2$ increases as soon as the absolute
value of the scattering length increases and one does not observe
any time delay in the compression of soliton trains
\cite{liang05,radha07}. It should also be mentioned that this time
delay in the compression of the soliton pulses can be suitably
manipulated by changing the trapping coefficient $\lambda(t)$.
Thus, the bright solitons can be compressed into a desired width
and amplitude in a controlled manner by suitably changing the trap
and our observation is consistent with the numerical results
\cite{xue05}.

\begin{figure}
\begin{center}
\includegraphics[width=0.65\linewidth]{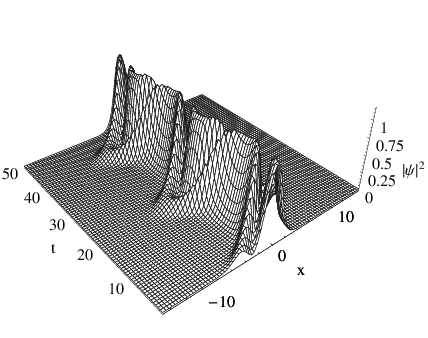}
\caption{Two soliton interaction in the confining trap
($\lambda(t)=0.09$) with $a(t)$ =$a_{0}$ $\exp(0.3 i t)$,
$a_{0}=0.5$, $\alpha_{10}=0.09$, $\beta_{10}=0.71$,
$\alpha_{20}=0.031$, $\beta_{20}=0.11$, $\phi_{1}= 5.1$,
$\delta_{1}=7.2$, $\phi_{2}=4.2$, $\delta_{2}=4.1$.}\label{td.4}
\end{center}
\end{figure}

It can also be observed that at $t>$14, the time dependent trap
becomes expulsive again for the same choice of $a(t)$
(\emph{{i.e.}}, $a(t) =0.5\exp(0.0025 t^2)$) (fig. \ref{td.3}b).
In order to sustain the confining nature of the trap
($\lambda(t)>$0), the scattering length should become complex as
it is being done in the case of cold alkaline earth metal atoms
\cite{ciury05}. Under this condition, the matter wave density
$|\psi|^2$ periodically changes with time by virtue of the
periodic modulation of scattering length \cite{kevre03,pelin03}
and this is reminiscent of the recent experimental observation of
Faraday waves \cite{engel07}. We also observe that the two soliton
pulses keep exchanging energy among themselves continuously during
propagation as shown in Fig. \ref{td.4}

We observe from the above that the addition of time dependence in
the trap enables one to stabilize the condensates for a longer
period of time by selectively tuning the trapping potential. It
should be mentioned that one can also selectively choose
$\lambda(t)$ and a(t) and observe their interplay in the
collisional dynamics of bright solitons. The interplay between
$\lambda(t)$ and a(t) consistent with equation (\ref{ch4-3eq11})
results in the ``matter-wave interference pattern" in the
collisional dynamics of bright solitons.

\begin{center}
\textbf{C.  Matter wave interference pattern in the collision of
bright solitons}
\end{center}

To generate the matter wave interference pattern, we now allow the
two bright solitons to collide with each other in the presence of
a trap for suitable choices of scattering length $a(t)$ and trap
frequency $\lambda(t)$ (or $c(t)$) consistent with equation
(\ref{ch4-3eq11}).

\begin{figure}
\begin{center}
\includegraphics[width=0.4\linewidth]{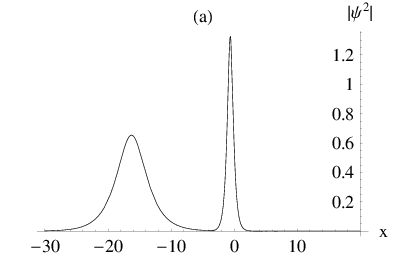}
\includegraphics[width=0.4\linewidth]{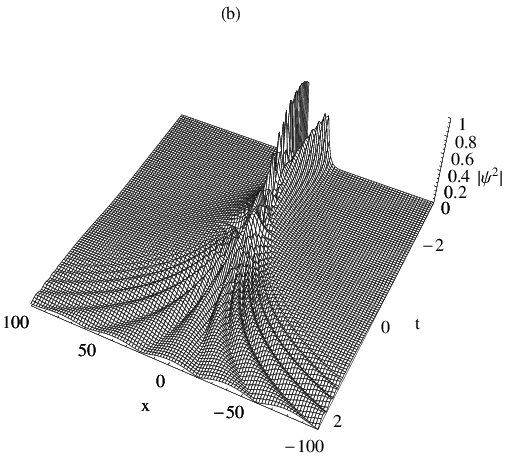}

\caption{(a) Initial position of two bright solitons at $t=-3$ for
$c(t)=-1$. (b) Interaction of two bright solitons forming the
interference pattern for the choice $a(t)= a_0 e^{-t}$ with
$\alpha_{10}$=0.01, $\alpha_{20}$=0.8, $\beta_{10}$=0.06,
$\beta_{20}$=0.012, $a_{0}$=0.2, $\phi_1$ =0.01, $\phi_2$=0.1,
$\delta_1$=0.1, $\delta_2$=0.01.}\label{inter1}
\end{center}
\end{figure}

\begin{figure}
\begin{center}
\includegraphics[width=0.4\linewidth]{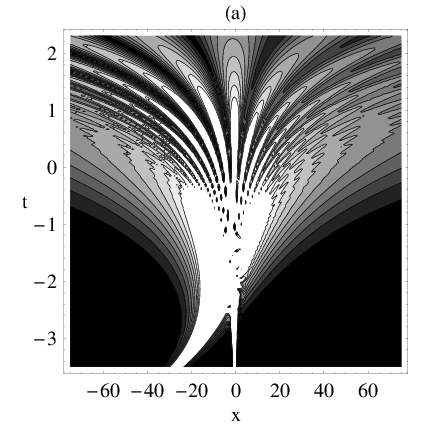}
\includegraphics[width=0.4\linewidth]{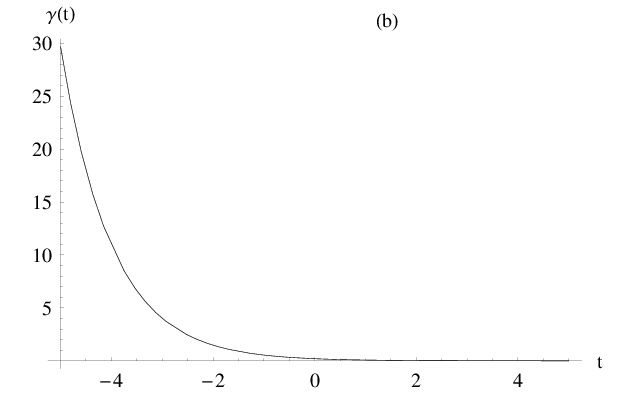}

\caption{(a)Contour plot of fig.\ref{inter1}(b) depicting matter
wave density evolution, (b) The time evolution of scattering
length $a(t)$ corresponding to case (i).}\label{inter2}
\end{center}
\end{figure}

\begin{figure}
\begin{center}
\includegraphics[width=0.4\linewidth]{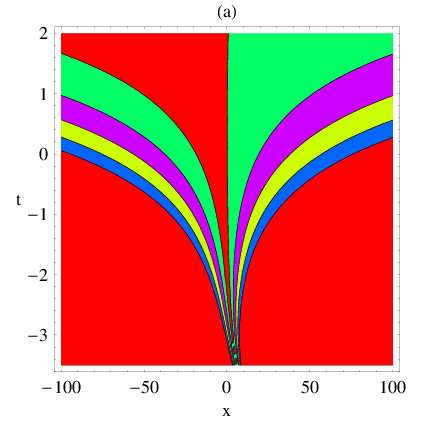}
\includegraphics[width=0.4\linewidth]{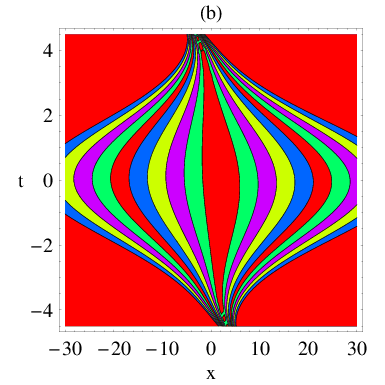}
\caption{(a) The phase difference between the condensates
corresponding to case (i) and (b) case (ii).}\label{inter4}
\end{center}
\end{figure}

\textbf{Case (i)}: When $c(t)=-1$, the scattering length evolves
as $a(t)=a_0 e^{-t}$ (shown in Fig. \ref{inter2}(b)) where $a_0$
is an arbitrary real constant and the trap frequency $\lambda(t)$
becomes expulsive and is equal to a constant ($\lambda(t)=-1$).
Under this condition, the collisional dynamics of two bright
solitons which are initially separated as shown in Fig.
\ref{inter1} (a) in the expulsive harmonic trap is shown in Fig.
\ref{inter1} (b) and the corresponding density evolution in Fig.
\ref{inter2} (a). The density evolution consists of alternating
bright and dark fringes of high and low density respectively while
the phase difference between the condensates continuously changes
with time as shown in Fig. \ref{inter4} (a).

\begin{figure}
\begin{center}
\includegraphics[width=0.35\linewidth]{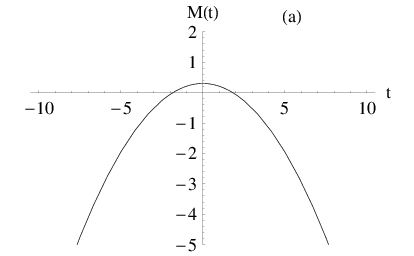}
\includegraphics[width=0.35\linewidth]{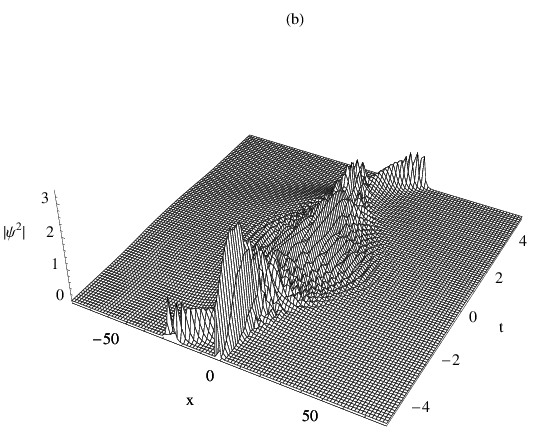}
\includegraphics[width=0.35\linewidth]{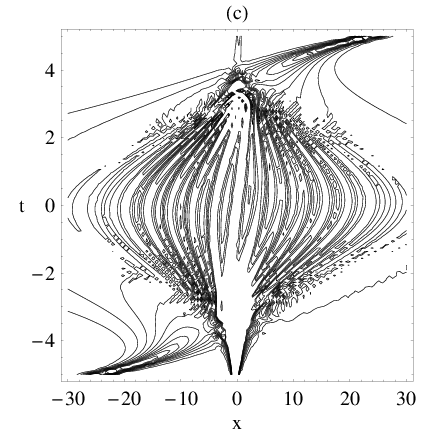}

\caption{(a) The time evolution of the trap $\lambda(t) =0.3-0.09
t^2$ corresponding to case (ii), (b) Two soliton interaction
forming the interference pattern corresponding to case (ii) for
the choice $a(t)=a_0 e^{(0.15t^2)}$ with $\alpha_{10}$=0.01,
$\alpha_{20}$=0.8, $\beta_{10}$=0.06, $\beta_{20}$=0.012,
$a_{0}$=0.02, $\phi_1$ =0.2, $\phi_2$=0.1, $\delta_1$=0.1,
$\delta_2$=0.2, (c) Contour plot showing the interference pattern
in the confining regime.}\label{inter3}
\end{center}
\end{figure}

\textbf{Case(ii)}: When $\lambda(t)=0.3-0.09 t^2$ (shown in Fig.
\ref{inter3}(a)) and $a(t) = a_0 e^{(0.15 t^2)}$, the soliton
interaction and the corresponding contour plot showing the
interference pattern in the confining region are shown in Figs.
\ref{inter3} (b) and \ref{inter3} (c) respectively. From the Figs.
\ref{inter3} (a)-(c), it is clear that when the trap $\lambda(t)$
enters the confining region from the expulsive domain, the
interaction of the solitons generates the interference pattern in
the confining regime and the pattern disappears once the trap
becomes expulsive again. The above choice of $\lambda(t)$ and
$a(t)$ can be synthesized easily under suitable laboratory
conditions. The phase difference between the condensates
continuously changes with time as shown in Fig. \ref{inter4} (b)

Thus, our results reinforce the fact that the matter waves
originating from the condensates (bright solitons) do interfere
and produce a fringe pattern analogous to the coherent laser beams
and the interference pattern is a clear signature of the long
range spatial coherence of the condensates. The interference
pattern generated by virtue of the collision of two bright
solitons \cite{rames09pla} is analogous to the interference
pattern obtained earlier experimentally by Andrews \emph{{et al.}}
\cite{andre97} wherein the two condensates were separated with a
sheet of green light and overlapped in ballistic expansion
(switching off the trap) while we selectively tune the frequency
of the trap $\lambda(t)$ in accordance with the scattering length
$a(t)$ consistent with equation (\ref{ch4-3eq11}). These
interference patterns are completely different from the
interference of two independent condensates originating from two
different traps with or without phase
\cite{javan96,javan86,javan91}.

It should be mentioned that the optical traps have opened up the
possibility of realizing different types of temporal variation of
the trap frequency $\lambda(t)$ while the scattering length can be
controlled both by Feshbach resonance as well as through the trap
frequency $\lambda(t)$. The phase change evolution shown in Figs.
\ref{inter4} (a) and \ref{inter4} (b) gives a measure of the
coherence of the condensates. It must be added that though the
concept of coherence of matter waves was already exploited to
create atom lasers, we do reconfirm the coherent nature of BECs in
the intra trap collision of bright solitons.

\begin{center}
\textbf{D. Dynamics of BECs with two and three body interactions}
\end{center}
In equation (\ref{ch4eq2}), when the scattering length
exponentially increases with time via Feshbach resoncance, the
density of the condensates also increases \cite{liang05,radha07}.
Naturally, a question arises as to how far one can increase the
scattering length to produce high density condensates. Since
collapse of the condensates sets in once the density exceeds a
critical value and one does not expect the collapse of BECs for a
true one-dimensional system, there is a constraint on increasing
the density of the condensates to sustain an effective true
one-dimensional BEC \cite{liang05}. This implies that one has to
investigate the dynamics of BECs within a safe range of
parameters. Hence, one has to look for an alternative to generate
high density condensates retaining the one-dimensionality of the
system without restricting to a parametric domain. It was observed
that for a large number of bosons, the repulsive three-body
interactions can overcome the two-body attractive interactions
thereby enhancing the stability of the condensates \cite{nref14}.

\begin{figure}
\begin{center}
\includegraphics[width=0.3\linewidth]{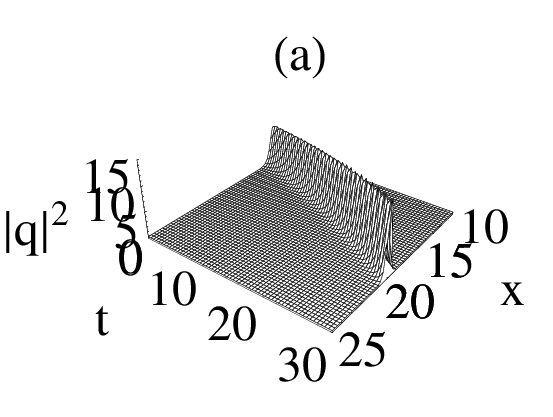}
\includegraphics[width=0.3\linewidth]{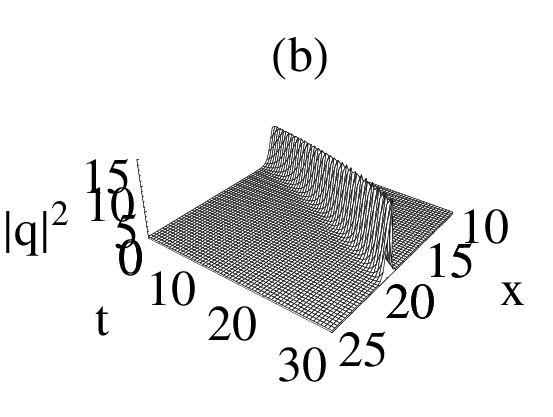}
\includegraphics[width=0.3\linewidth]{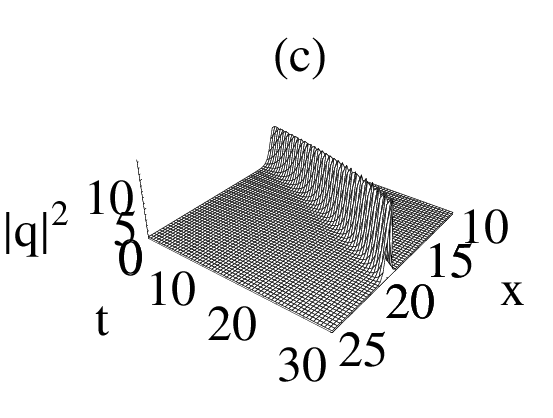}
\caption{Density of condensates in the modified GP equation with
both attractive two-body and attractive three-body interactions
for the parametric choice $\lambda$=0.02, $\alpha_0$=0.1,
$\beta_0$=0.9, $\delta_1$=$-$1.5, $\phi_1$=2.5 for (a)
$\tilde{a}_0$=0.4, $\tau=0$, (b)$\tilde{a}_0$=0.4, $\tau=0.04$,
(c)$\tilde{a}_0$=0.45, $\tau=0.04$.}\label{ttf1}
\end{center}
\end{figure} From the above, we understand that two-body interactions
alone are insufficient and they have to be suitably combined with
three-body interactions to extend the region of stability while
increasing the density of the condensates in a quasi
one-dimensional regime.

\begin{figure}
\begin{center}
\includegraphics[width=0.4\linewidth]{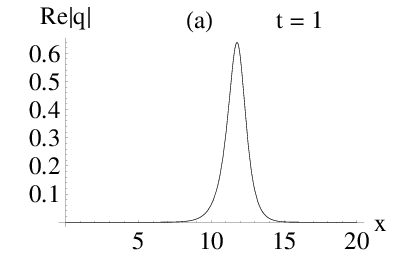}
\includegraphics[width=0.4\linewidth]{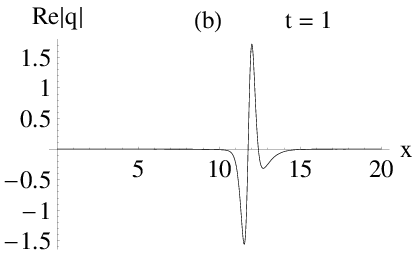}
\caption{Real part of the order parameter of (a) cubic GP equation
and (b) modified GP equation with attractive three-body
interactions. }\label{ttf2}
\end{center}
\end{figure}
\begin{figure}
\begin{center}
\includegraphics[width=0.4\linewidth]{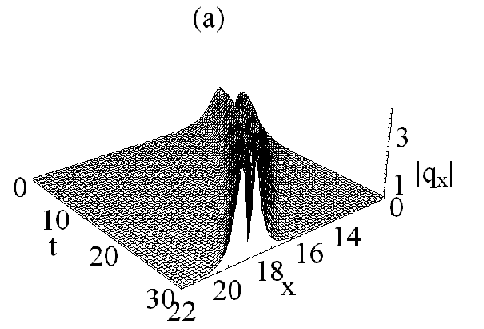}
\includegraphics[width=0.4\linewidth]{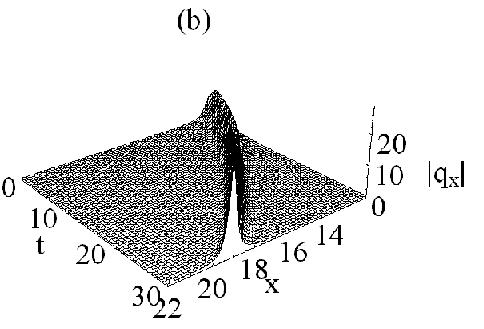}
 \caption{(a)
Compression and splitting of $|q_x|$ of cubic GP equation (b)
Compression and suppression of splitting of $|q_x|$ in the
modified GP equation.}\label{ttf3}
\end{center}
\end{figure}

\begin{figure}
\begin{center}
\includegraphics[width=0.3\linewidth]{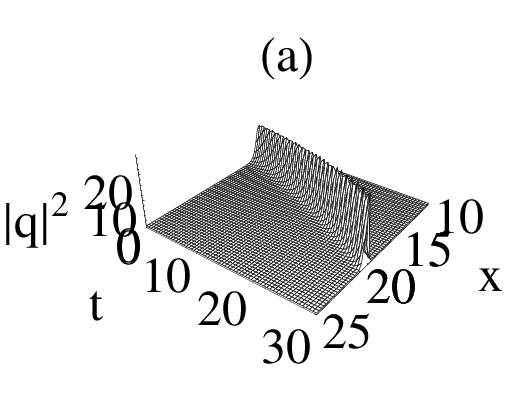}
\includegraphics[width=0.3\linewidth]{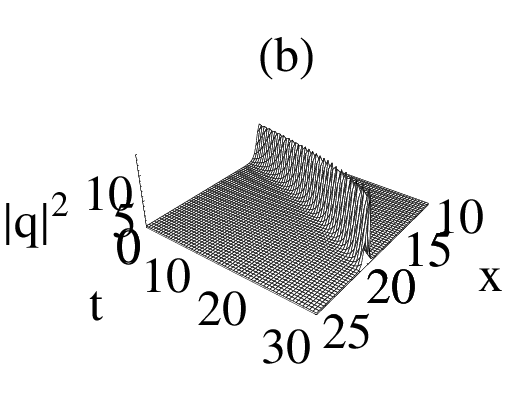}
\includegraphics[width=0.3\linewidth]{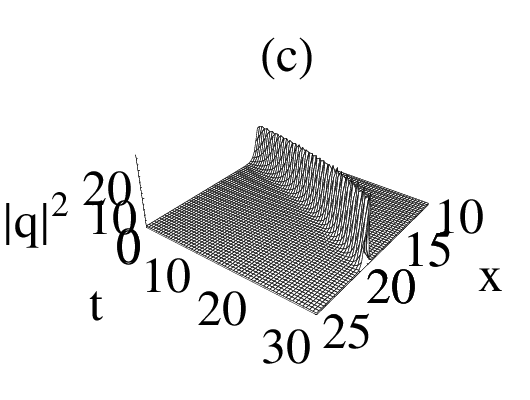}
\caption{Density of condensates in the modified GP equation with
attractive two-body interactions and repulsive three-body
interactions with the same parametric choice as in fig.\ref{ttf1}
for (a) $\tilde{a}_0$=0.6, $\bar{\tau}=0.04$,
(b)$\tilde{a}_0$=0.8, $\bar{\tau}=0.04$, (c)$\tilde{a}_0$=0.8,
$\bar{\tau}=0.09$.}\label{ttf4}
\end{center}
\end{figure}

\begin{figure}
\begin{center}
\includegraphics[width=0.4\linewidth]{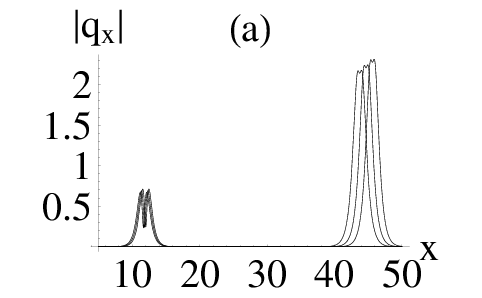}
\includegraphics[width=0.4\linewidth]{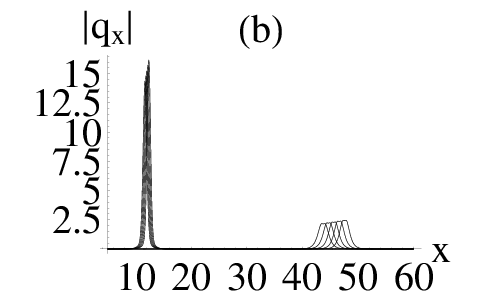}
\caption{ Asymptotic forms of the density profiles $|q_x|$ of the
two-soliton solution of (a) cubic GP equation and (b) modified GP
equation (attractive).}\label{ttf5}
\end{center}
\end{figure}

To introduce the new integrable model describing the impact of
both two- and three-body interactions on the condensates, we now
consider an additional phase imprint on the order parameter
$\psi(x,t)$ to generate a new order parameter $q(x,t)$ as
\begin{equation}
q(x,t) = \psi(x,t) \rm{e}^{2 i \theta(x,t)},
\end{equation}
where $\theta(x,t)$ is the phase-imprint on the old order
parameter $\psi(x,t)$. We now engineer the phase imprint
$\theta(x,t)$ in accordance with the following equations;
\begin{eqnarray}
\theta_x = -\sqrt{\tau} |\psi|^2, \label{phaserelation1}\\
\theta_t = i \sqrt{\tau} (\psi \psi^*_x - \psi^* \psi_x)+4 \tau
|\psi|^4,\label{phaserelation2}
\end{eqnarray}
so that the transformed order parameter $q(x,t)$ obeys an
evolution equation
\begin{equation}
i q_t + q_{xx} + 2 a(t)|q|^2 q+\frac{\lambda^2 x^2}{4}
q+4\tau|q|^4 q+4i\sqrt{\tau}(|q|^2)_x q =0.\label{ttint}
\end{equation}

In equation(\ref{ttint}), $a(t)(=\tilde{a}_0$exp$(\lambda t)$)
represents attractive ($\tilde{a}_0>0$) two-body interactions
while the real and arbitrary parameter $\tau$ corresponds to the
strength of three-body interactions assuming that the contribution
of three-atom collisions to the loss is negligible (by kicking out
atoms from the condensate into thermal cloud)
\cite{nref14}-\cite{nref19}. This type of engineering the phase
imprint to generate a new integrable model describing the impact
of both two- and three-body interactions on the condensates is
reminiscent of generating solitons by phase engineering of BECs of
sodium and rubidium \cite{nref22}.

Making use of the gauge transformation approach, we obtain the
bright one soliton solution of the modified GP equation
(\ref{ttint})\cite{jpsj}

\begin{equation}
q^{(1)}=\frac{2}{\sqrt {\tilde{a}_0}} \beta_{0} \rm{e}^{\left(
\frac{{\lambda t}}{2} - \frac{{i\lambda x^2 }}{4}+2i\theta \right
)}\rm{sech}(\chi_1)\rm{exp}(i\xi_{1}),\label{ttsol}
\end{equation}

\begin{eqnarray}
\chi_1&=&2\beta_{1}x -8\int (\alpha_{1}\beta_{1})dt+2\delta_1,\label{ttsolchi}\\
\xi_1&=&2\alpha_{1}x -4\int(\alpha_{1}^2 - \beta_{1}^2)dt-2\phi_1,\label{ttsolxi}\\
\alpha_1&=&\alpha_{0}\rm{e}^{\lambda
t},\;\;\;\;\beta_1=\beta_{0}\rm{e}^{\lambda t},\label{ttsolalpha}
\end{eqnarray}
and
\begin{equation}
\theta=-\frac{2}{\tilde{a}_0} \sqrt{\tau} \beta_0
\rm{tanh}\left[2\left(\frac{\rm{e}^{\lambda t}(x \lambda - 2
\rm{e}^{\lambda t} \alpha_0)\beta_0}{\lambda}+\delta_1\right)
\right],\label{phaseterm}
\end{equation}
where $\alpha_0$, $\beta_0$, $\delta_1$ and $\phi_1$ are arbitrary
real constants. It should be mentioned that the phase imprint
$\theta$ given by equation (\ref{phaseterm}) is related to the
density of the old order parameter $\psi$ (by virtue of equations
(\ref{phaserelation1}) and (\ref{phaserelation2}) and hence the
bright solitons given by equation(\ref{ttsol}) are endowed with
phase dependent amplitude. It should also be added that equation
(\ref{ttint}) admits bright solitons only for attractive two-body
interactions ($\tilde{a}_0>0$) while the three-body interactions
can be either attractive ($\tau >0$) or repulsive ($\tau<0$).

\textbf{I. Condensates with attractive three-body interactions
($\tau > 0$)}

From the equations (\ref{ttsol})-(\ref{phaseterm}), one observes
that the bright solitons of the integrable modified GP equation
(\ref{ttint}) acquires a kink like additional nontrivial phase
represented by equation (\ref{phaseterm}) in comparison with that
of cubic GP solitons ($\tau=0$). Figure \ref{ttf1} (a) portrays
the density evolution of the condensates in the absence of
three-body interactions. When one considers the attractive
three-body interactions in addition to the attractive two-body
interactions with the strength of both being equal, one does not
observe any perceptible change in the matter wave density as shown
in Fig. \ref{ttf1} (b). When the strength of attractive two-body
interactions is increased, one observes a decrease in the density
of condensates as shown in Fig. \ref{ttf1} (c). From this, one
understands that an instability sets in the condensates leading to
the ejection of the atoms resulting in the decrease of matter wave
density. Hence, to ensure the stability of the condensates over a
longer interval of time, the strength of the attractive two-body
interactions should be minimum.

It is also obvious that this additional phase arising in the
dynamics of the condensates of the modified GP equation with
attractive three-body interactions evolves in space and time as
indicated by equation (\ref{phaseterm}) and hence one understands
that its effect will be more pronounced either in the real (or
imaginary) part or in the derivatives of the order parameter.
Figure \ref{ttf2} (b) shows the effect of the nontrivial phase on
the matter wave solitons of the modified GP equation in comparison
with that of cubic GP equation shown in Fig. \ref{ttf2} (a). Thus,
it is evident that the additional nontrivial phase contributes to
the compression and rarefaction of the matter wave solitons of the
modified GP equation.

From the spatial derivative $|q_x|$ plotted in Figs. \ref{ttf3}
(a) and \ref{ttf3} (b), one infers that as the atoms start
accumulating in the lowest quantum state, the matter wave density
increases leading to a compression of the matter wave of the cubic
GP equation. In addition to the compression, there is a splitting
of the matter wave in the cubic GP equation (see Fig. \ref{ttf3}
(a)), while this splitting of the matter wave has been completely
suppressed by the attractive three-body interactions as shown in
Fig. \ref{ttf3} (b).

\textbf{II. Condensates with repulsive three-body interactions
($\tau<0$)}

For repulsive three-body interactions, the bright soliton of
integrable modified  GP equation becomes
\begin{equation}
\bar{q}^{(1)}=\frac{2}{\sqrt
{\tilde{a}_0}}\beta_{0}\rm{sech}(\chi_1)\rm{exp}(i\xi_{1}) \rm
{e}^{\left( \frac{{\lambda t}}{2} - \frac{{i\lambda x^2
}}{4}+\bar{\theta}\right)},\label{ttsol2}
\end{equation}
where $\bar{\theta}=\frac{4}{\tilde{a}_0}\sqrt{\bar{\tau}} \beta_0
\rm{tanh}\left[2\left(\frac{\rm{e}^{\lambda t}(x \lambda - 2
\rm{e}^{\lambda t}
\alpha_0)\beta_0}{\lambda}+\delta_1\right)\right]$ with
$\bar{\tau}$ being a real parameter.

$\quad$Figure \ref{ttf4} (a) displays the density evolution of the
condensates for attractive two-body interactions and repulsive
three-body interactions. Thus, the addition of a small repulsive
three-body force contributes to an exponential increase in the
matter wave density (as the amplitude of the condensates becomes
$\frac{2}{\sqrt{\tilde{a}_0}}\beta_0 \rm{exp}(\lambda t/2
+\bar{\theta})$ from equation (\ref{ttsol2}) in comparison to the
density evolution of the condensates in the absence of three-body
interactions shown in Fig .\ref{ttf1}(a). The fact that the
condensates can hold enormous number of atoms together with a
small repulsive force means that one can extend the region of
stability of the condensates in a quasi one-dimensional regime.
When one increases the strength of attractive two-body
interactions, the density of the condensates decreases as shown in
Fig. \ref{ttf4} (b) as compared to Fig. \ref{ttf4} (a), thereby
setting in an instability in the condensates. However, this
instability can be overcome by increasing the strength of
repulsive three-body interactions as shown in Fig. \ref{ttf4} (c),
wherein the matter wave density again increases enormously in
comparison with Fig. \ref{ttf4} (b) and attains the magnitude
shown in Fig. \ref{ttf4} (a). Thus, we observe that the addition
of repulsive three-body interactions can allow the number of
condensed atoms to increase enormously and this can happen even
when the strength of the repulsive three-body interactions is very
small compared with the strength of two-body
interactions\cite{nref14}. It should be mentioned that eventhough
the bright solitons of the modified GP equation with both two- and
three-body interactions (both repulsive and attractive) in the
expulsive potential are unstable again, the extent of instability
has reduced compared to that of a cubic GP equation with two-body
interactions alone.

\section{Dynamics of Vector Bose-Einstein Condensates} We know that the behaviour of
single component (or scalar) BECs is influenced by the external
trapping potential and interatomic interaction. Experimental
realization of BECs in which two (or more) internal states or
different atoms can be populated has given a fillip to the
investigation of multicomponent BECs \cite{ho96,esry97,pu98}. In
contrast to the single component BECs, the behaviour of
multicomponent condensates is much richer because of both intra
species interaction and inter species interaction. This extra
freedom arising by virtue of the interaction among the internal
states or different atoms offers multicomponent BECs several
interesting and complicated properties which are not witnessed in
single component condensates. Far from being a trivial extension
of the single component BECs, multicomponent condensates exhibit
novel and rich phenomena such as soliton trains, soliton pairs,
multidomain walls, spin-switching \cite{ieda04,uchiy06} and
multimode collective excitations.

Motivated by the above considerations, we investigate the dynamics
of a two-component BEC in a time dependent harmonic trap described
by a two-coupled GP equation and deduce the integrability
condition for the existence of vector bright solitons. We then
bring out the fascinating collision of vector bright solitons
demonstrating the switching of energy underscoring the longevity
of vector BECs. We then employ Feshbach resonance management to
manoeuvre the scattering length to further enhance the lifetime of
two component BECs.
\begin{center}
\textbf{A.  Mathematical Model}
\end{center}

Considering a two-component BEC, the behaviour of the condensates
that are prepared in two hyperfine states can be described at
sufficiently low temperatures by the two coupled GP equation of
the following form \cite{ho96,esry97,pu98}
\begin{eqnarray}
i\hbar \frac{\partial \psi_1}{\partial t} =
\left(-\frac{\hbar^2}{2
m_1}\nabla^2+U_{11}|\psi_1|^2+U_{12}|\psi_2|^2
+V_1 \right)\psi_1,\label{ch6eq2}\\
i\hbar \frac{\partial \psi_2}{\partial t} =
\left(-\frac{\hbar^2}{2
m_2}\nabla^2+U_{21}|\psi_1|^2+U_{22}|\psi_2|^2 +V_2
\right)\psi_2,\label{ch6eq3}
\end{eqnarray}
where the condensate wave functions are normalized through the
particle numbers $N_i = \int |\psi_i|^2 d^3 {\textbf{r}}$.
$U_{ii}=4 \pi \hbar^2 a_{ii}/m$ and $U_{ij} = 2\pi \hbar^2
a_{ij}/m$ represent intraspecies and interspecies interaction
strengths respectively with $a_{ij}$ being the corresponding
scattering lengths and $m$ is the reduced mass. The trapping
potentials are assumed to be $V_i = m_i[\omega_{ix}^2 x^2
+\omega_{i\bot}^2 (y^2+z^2)]/2$. Further assuming that $\omega_{i
\bot} \gg \omega_{ix}$ such that the transverse motions of the
condensates are frozen to the ground state of the transverse
harmonic trapping potential, the system becomes quasi one
dimensional in nature. Integrating out the transverse coordinates,
the resulting equations for the axial wave functions
$\psi_{1,2}(x,t)$ in dimensionless form can be written as
\begin{eqnarray}
i\frac{\partial \psi_1}{\partial t}=
\left(-\frac{1}{2}\frac{\partial^2}{\partial x^2}+b_{11}|\psi_1|^2
+b_{12}|\psi_2|^2 +\frac{\lambda_1^2}{2}x^2 \right)\psi_1,\label{ch6eq4}\\
i\frac{\partial \psi_2}{\partial t}=
\left(-\frac{k}{2}\frac{\partial^2}{\partial x^2}+b_{21}|\psi_1|^2
+b_{22}|\psi_2|^2+\frac{\lambda_2^2}{2 k }x^2
\right)\psi_2,\label{ch6eq5}
\end{eqnarray}
where units for length and time become $\sqrt{\frac{\hbar}{m_1
\omega_{1\bot}}}$ and $2\pi/\omega_{1 \bot}$ and $\psi_{1,2}$ is
normalized such that $\int|\psi_1|^2 dx =1$ and $\int |\psi_2|^2
dx =N_2/N_1$. Other parameters are defined as: $b_{11} = 2 a_{11}
N_1$, $b_{12} = 2m_1
a_{12}N_1/[(1+\omega_{2\bot}/\omega_{1\bot})m]$, $b_{21} =2m_1
a_{21}N_1/[(1+\omega_{2\bot}/\omega_{1\bot})m]$, $b_{22}=2 a_{22}k
N_1 \omega_{2\bot}/\omega_{1\bot}$, $\lambda_1 =
\omega_{1x}/\omega_{1\bot}$,
$\lambda_2=\omega_{2x}/\omega_{1\bot}$ and $k=m_1/m_2$.

Now, we consider $k=1$ (\emph{i.e.}, $m_1 = m_2$) and $\omega_1 =
\omega_2 = \omega$ (\emph{i.e.}, $\lambda_1 = \lambda_2$) and
allow the scattering lengths $a_{ij}$ and the strength of trapping
potential $\lambda^2 (\lambda= \omega_x /\omega_\bot)$ to vary
with time. Then, the above equation with $\tilde{t} = t/2$ takes
the following form (after omitting tilde)

\begin{eqnarray}
i\psi_{1t}+\psi_{1xx}+2(b_{11}(t)|\psi_1|^2+b_{12}(t)|\psi_2|^2)\psi_1
+\lambda(t)^2 x^2 \psi_1 = 0,\label{eq:gp:general1} \\
i \psi_{2t}+\psi_{2xx}+2(b_{21}(t)|\psi_1|^2
+b_{22}(t)|\psi_2|^2)\psi_2+\lambda(t)^2 x^2 \psi_2 =
0,\label{eq:gp:general2}
\end{eqnarray}
In the above equation, $b_{ij} (i,j=1,2)$ represents the
attractive interaction strength and the trap frequency could be
both confining ($\lambda(t)^2>0$) and expulsive ($\lambda(t)^2 <
0$). When the longitudinal trapping potential is neglected
($\lambda(t)$=0) and $b_{11}=b_{12}=b_{21}=b_{22}=\rm {a~
constant}$, the system becomes the celebrated Manakov model (see
Refs. \cite{makha81} to \cite{liu09}) admitting shape changing
collision of vector solitons.

When both intraspecies interaction and interspecies interaction
are all equal and time dependent (\emph{i.e.}, $b_{11}(t)=
b_{12}(t)=b_{21}(t)=b_{22}(t)=a(t)$), the above equations
(\ref{eq:gp:general1}) and (\ref{eq:gp:general2}) take the
following form

\begin{eqnarray}
i\psi_{1t}+\psi_{1xx}+2a(t)(|\psi_1|^2+|\psi_2|^2)\psi_1 + \lambda(t)^2 x^2 \psi_1 = 0, \label{eq:gp:symmetric1} \\
i \psi_{2t}+\psi_{2xx}+2a(t)(|\psi_1|^2
+|\psi_2|^2)\psi_2+\lambda(t)^2 x^2 \psi_2 = 0,
\label{eq:gp:symmetric2}
\end{eqnarray}
The above coupled GP equation has been investigated and shown to
admit integrability for symmetric interaction strengths
\cite{liu09,zhang09,rajen09}.
\begin{center}
\textbf{B.  Lax-pair and integrability condition}
\end{center}

Equations (\ref{eq:gp:symmetric1}) and (\ref{eq:gp:symmetric2})
admit the following linear eigenvalue problem

\begin{eqnarray}
\Phi_x &+& U \Phi=0,\\
\Phi_t &+& V \Phi=0,
\end{eqnarray}\label{lax}
where $\Phi = (\phi_1, \phi_2, \phi_3)^T$ and
\begin{eqnarray}
U &=& \left(%
\begin{array}{ccc}
i\zeta(t) & Q_{1} &  Q_{2}\\
Q_{1}^*& -i\zeta(t) & 0 \\
Q_{2}^*& 0 & -i\zeta(t) \\
\end{array}%
\right),
\end{eqnarray}
\begin{eqnarray}
V&=&\left(%
\begin{array}{ccc}
V_{11} & V_{12}& V_{13} \\
V_{21} & V_{22} & V_{23} \\
V_{31} & V_{32} & V_{33} \\
\end{array}%
\right),\notag\\
\end{eqnarray}
with
\begin{eqnarray}
v_{11}&=& -i \zeta(t)^{2}+i \Gamma(t) x \zeta(t) + \frac{i}{2} Q_{1} Q_{1}^*+ \frac{i}{2} Q_{2}Q_{2}^*\nonumber\\
v_{12}&=& (\Gamma(t) x- \zeta(t)) Q_1 + \frac{i}{2} Q_{1x}\nonumber\\
v_{13}&=& (\Gamma(t) x- \zeta(t)) Q_2 + \frac{i}{2} Q_{2x}\nonumber\\
v_{21}&=& -(\Gamma(t) x- \zeta(t)) Q_1^* + \frac{i}{2} Q_{1x}^*\nonumber\\
v_{22}&=& i \zeta(t)^{2}-i \Gamma(t) x \zeta(t) - \frac{i}{2} Q_{1} Q_{1}^* \nonumber\\
v_{23}&=& - \frac{i}{2} Q_{2} Q_{1}^* \nonumber\\
v_{31}&=& -(\Gamma(t) x- \zeta(t)) Q_2^* + \frac{i}{2} Q_{1x}^*\nonumber\\
v_{32}&=& - \frac{i}{2} Q_{1} Q_{2}^* \nonumber\\
v_{33}&=& i \zeta(t)^{2}- i \Gamma(t) x \zeta(t) - \frac{i}{2}
Q_{2}Q_{2}^*\nonumber
\end{eqnarray}
\begin{eqnarray}
Q_1=\frac{1}{\sqrt{A(t)}} \psi_1(x,t) e^{i \Gamma(t) x^2} \notag\\
Q_2=\frac{1}{\sqrt{A(t)}} \psi_2(x,t) e^{i \Gamma(t) x^2} \notag
\end{eqnarray}
It is obvious that the compatibility condition
$(\Phi_x)_t$=$(\Phi_t)_x$ leads to the zero curvature equation
$U_t-V_x+[U,V]=0$ which yields the integrable coupled GP equation
(\ref{eq:gp:symmetric1}) and (\ref{eq:gp:symmetric2}) provided the
spectral parameter $\zeta(t)$ obeys the following nonisospectral
condition
\begin{equation}
\zeta(t)=\mu e^{-(\int \Gamma(t) dt)}\label{eq:mu}
\end{equation}
where $\mu$ is a hidden complex constant and $\Gamma(t)$ is an
arbitrary function of time and is related to the trap frequency

\begin{equation}
\lambda ^{2}(t)=\Gamma^{2}(t)-\Gamma^{\prime }(t).  \label{trap}
\end{equation}
Further, equation (\ref{trap}) which represents the trap frequency
$\lambda(t)$ can be related to the scattering length $a(t)$
through the integrability condition
\begin{equation}
-a^{\prime \prime }(t)a(t)+2 a^{\prime}(t) ^{2}-\lambda
^{2}(t)a^{2}(t)=0. \label{integ}
\end{equation}
It should be mentioned that the coupled GP equation represented by
(\ref{eq:gp:symmetric1}) and (\ref{eq:gp:symmetric2}) is
completely integrable for suitable choices of the trap frequency
$\lambda(t)$ and scattering length $a(t)$ consistent with equation
(\ref{integ}). For the constant trapping frequency, $\lambda
(t)=c_{1}$, equation (\ref{integ}) yields $a(t)=e^{c_{1}t}$.
\begin{figure}
\begin{center}
\includegraphics[width=0.45\linewidth]{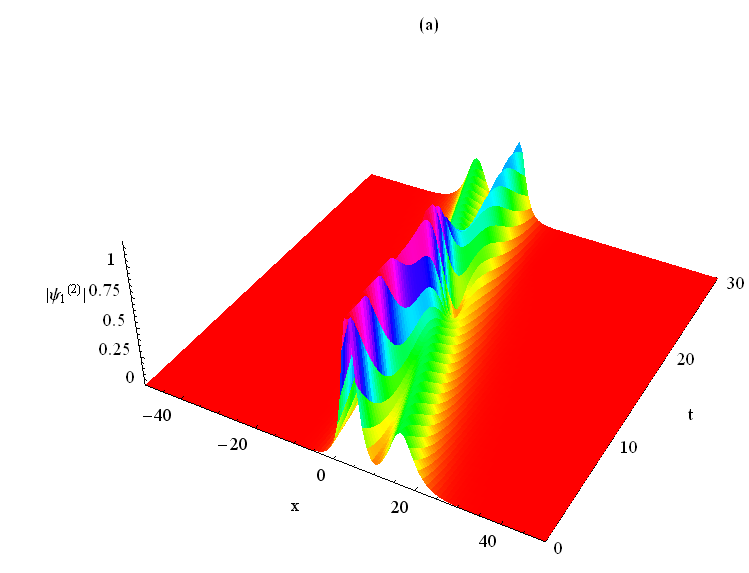}
\includegraphics[width=0.45\linewidth]{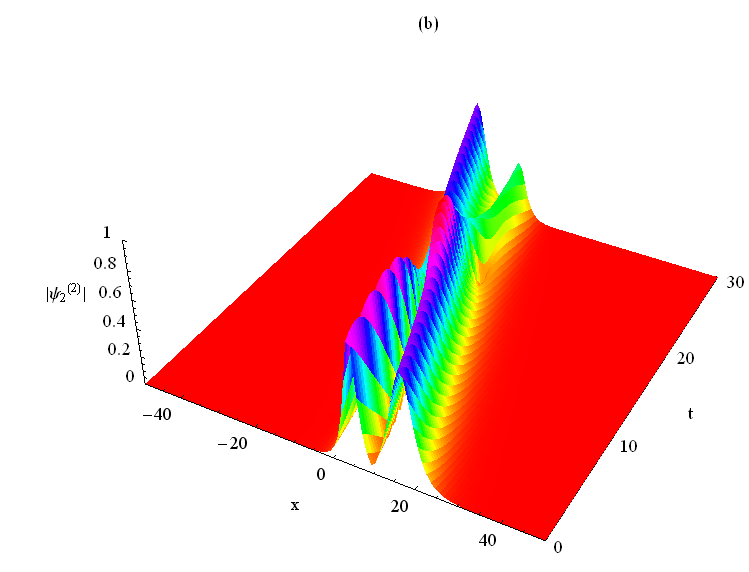}
\caption{Switching of matter wave (bright) solitons in two
component BECs for $a(t)$=0.01 and $\lambda(t)$= 0.01+0.0001
$t^2$}\label{twocompie}
\end{center}
\end{figure}
The bright solitons of equations
(\ref{eq:gp:symmetric1})-(\ref{eq:gp:symmetric2}) employing gauge
transformation method have the following form
\begin{eqnarray}
\psi_1^{(1)} = \sqrt{\frac{1}{a(t)}}\varepsilon_1^{(1)} 2
\beta_1(t) sech(\theta_1)e^{i(-\xi_1 + \Gamma(t) \frac{x^2}{2})},\label{coupledgponesol1}\\
\psi_2^{(1)} = \sqrt{\frac{1}{a(t)}}\varepsilon_2^{(1)} 2
\beta_1(t) sech(\theta_1)e^{i(-\xi_1 + \Gamma(t)
\frac{x^2}{2})},\label{coupledgponesol2}
\end{eqnarray}
where
\begin{eqnarray}
\theta_1 &=& 4\int \alpha_1 \beta_1 dt+2 x \beta_1-2 \delta_1,\\
\xi_1 &=& 2\int(\alpha_1^2-\beta_1^2)dt+2x\alpha_1-2\chi_1,
\end{eqnarray}
with $\alpha_1 = \alpha_{10} e^{[\int{\Gamma(t)}dt]} $,
$\beta_1=\beta_{10} e^{[\int{\Gamma(t)}dt]} $ while $\delta_1$ and
$\chi_1$ are arbitrary parameters. Gauge transformation approach
can  also be extended to generate multi soliton solution
\cite{twosol.vr.2010}. Figure \ref{twocompie} shows that it should
be possible to transfer (switch) energy from one mode to the other
in a vector BEC thereby one can enhance the longevity (or
lifetime) of the bright solitons (or the condensates). This can
happen irrespective of whether the trap is time dependent or
independent.

\begin{center}
\textbf{C.  Feshbach Resonance Management of vector bright
solitons}\cite{mypapertobe}
\end{center}

It is obvious from the investigation of vector BECs that the
lifetime of the condensates gets enhanced by virtue of the
switching of energy between the two components. It should be
mentioned that Feshbach resonance management can also be employed
to drive home the fact that vector BECs in a time dependent
harmonic trap are longlived compared to the condensates dwelling
in a time independent trap.

\begin{figure}
\begin{center}
\includegraphics[scale=0.25]{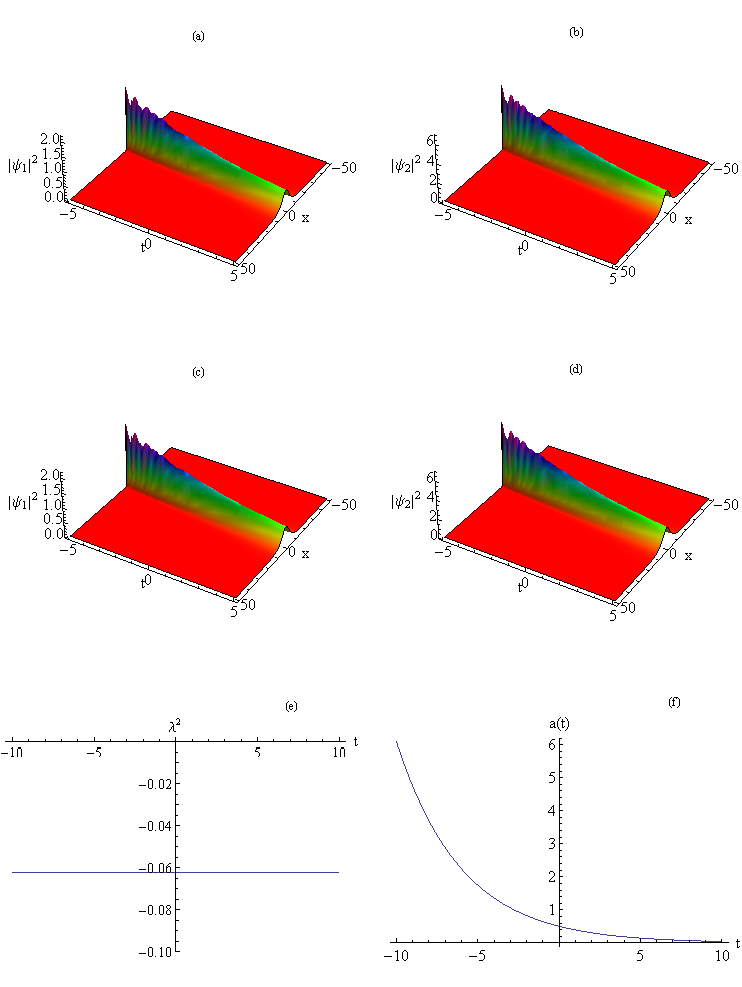}
\caption{{\bf Upper Panel (a)-(b)}: Density (Analytical) profile
of the condensates (bright solitons) in the time independent
expulsive trap for $\Gamma(t)$= -0.25 and $a(t)=0.5e^{-0.25 t}$;
{\bf Middle Panel (c)-(d)}: Numerically simulated density profile
for $\Gamma(t)$= -0.25 and $a(t)=0.5e^{-0.25 t}$; {\bf Lower
panel(e)-(f)}: Trap strength and binary interaction
}\label{fig.ana1}
\end{center}
\end{figure}

\begin{figure}
\begin{center}
\includegraphics[scale=0.25]{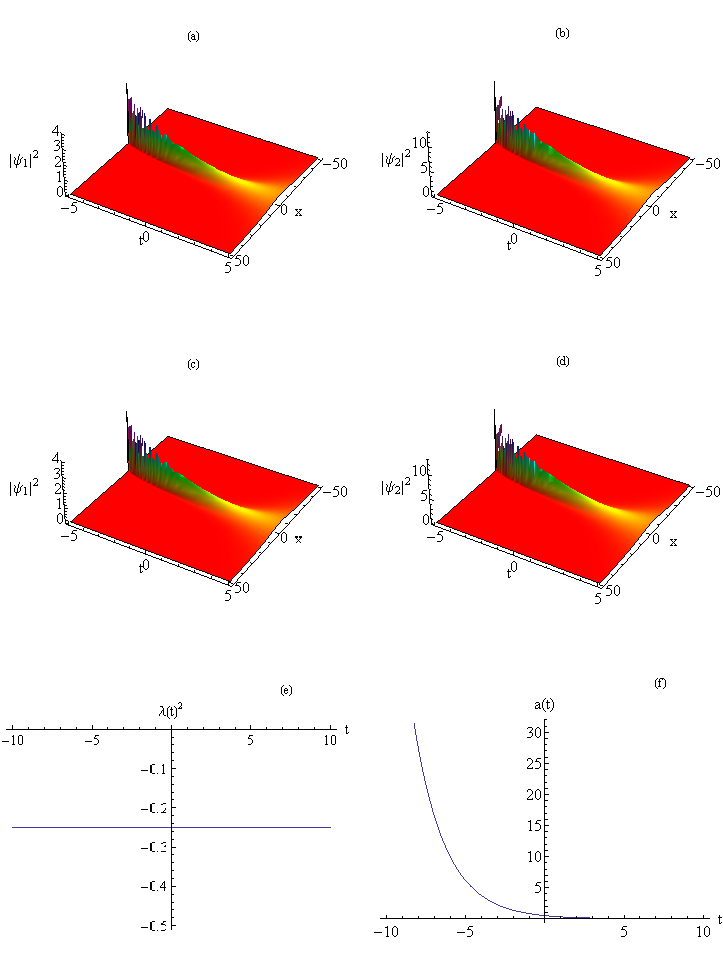}
\caption{{\bf Upper panel (a)-(b)}: Compression (Analytical) of
the condensates in the time independent expulsive trap for
$\Gamma(t)$= -0.5 and $a(t)=0.5e^{-0.5 t}$; {\bf Middle Panel
(c)-(d)}: Numerically simulated density profile for $\Gamma(t)$=
-0.5 and $a(t)=0.5e^{-0.5 t}$ showing the compression of BECs;
{\bf Lower panel (e)-(f)}: Trap strength and binary interaction.
}\label{fig.ana2}
\end{center}
\end{figure}

\begin{figure}
\begin{center}
\includegraphics[scale=0.25]{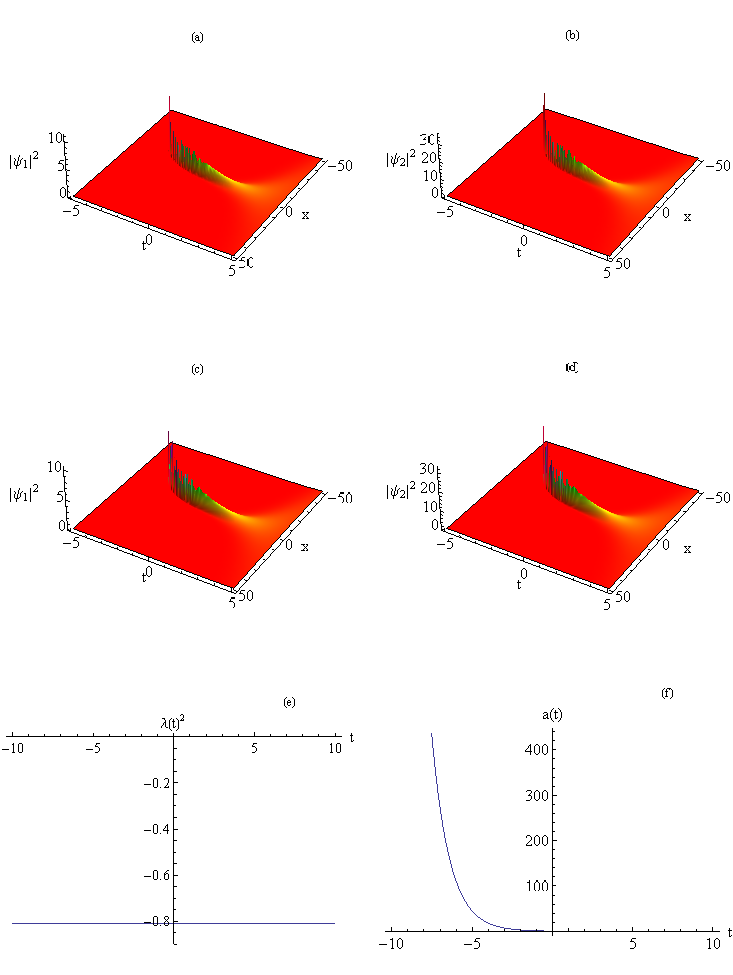}
\caption{{\bf Upper panel (a)-(b)}: Onset of collapse (Analytical)
of the condensates for $\Gamma(t)$= -0.9 and  $a(t)=0.5e^{-0.9 t}$
in an expulsive time independent trap; {\bf Middle Panel (c)-(d)}:
Numerically simulated density profile for $\Gamma(t)$= -0.9 and
$a(t)=0.5e^{-0.9 t}$ showing the collapse of BECs; {\bf Lower
panel (e)-(f)}: Trap strength and binary
interaction}\label{fig.ana3}
\end{center}
\end{figure}

To start with, we now switch off the time dependence of the
harmonic trap and keep track of the evolution of the condensates
by manipulating the scattering length of the form $a(t)=0.5
e^{-0.25 t}$ (shown in Fig. \ref{fig.ana1} (f)) consistent with
equations (\ref{trap}) and (\ref{integ}), thereby rendering
($\Gamma(t)= -0.25$) the trap expulsive (shown in Fig.
\ref{fig.ana1} (e)). The corresponding density profile of the
condensates is shown in the upper panel of Fig. \ref{fig.ana1}
(Figs. \ref{fig.ana1} (a) and \ref{fig.ana1} (b)). The associated
numerically simulated density profile of the condensates employing
real time propagation of Split Step Crank Nicolson method is shown
in Figs. \ref{fig.ana1} (c) and \ref{fig.ana1} (d). From Figs.
\ref{fig.ana1} (a)-\ref{fig.ana1} (d), we observe that there is a
perfect agreement between analytical and numerical results. When
we double the trap frequency ($\Gamma(t)= -0.5$) keeping the trap
expulsive again (shown in Fig. \ref{fig.ana2} (e)) and manipulate
the scattering length through Feshbach resonance of the form
$a(t)=0.5 e^{-0.5 t}$ (shown in Fig. \ref{fig.ana2} (f))
consistent with equations(\ref{trap}) and (\ref{integ}), the
compression (analytical) the condensates sets in the two modes as
shown in Figs. \ref{fig.ana2} (a)-(b). This is again confirmed by
the numerical simulation shown in Figs. \ref{fig.ana2} (c) and
\ref{fig.ana2} (d). When we further enhance the trap strength (3.6
times the original strength) keeping the trap expulsive again
(shown in Fig. \ref{fig.ana3} (e)) and further manipulate
scattering length a(t) of the form $a(t)=0.5 e^{-0.9 t}$ (shown in
Fig. \ref{fig.ana3}(f)), one observes the onset of collapse of the
condensates as shown in Figs. \ref{fig.ana3} (a)-(b). Again, the
numerically simulated condensates shown in Figs. \ref{fig.ana3}
(c) and \ref{fig.ana3} (d) match with Figs. \ref{fig.ana3} (a) and
\ref{fig.ana3} (b) respectively.

\begin{figure}
\begin{center}
\includegraphics[scale=0.25]{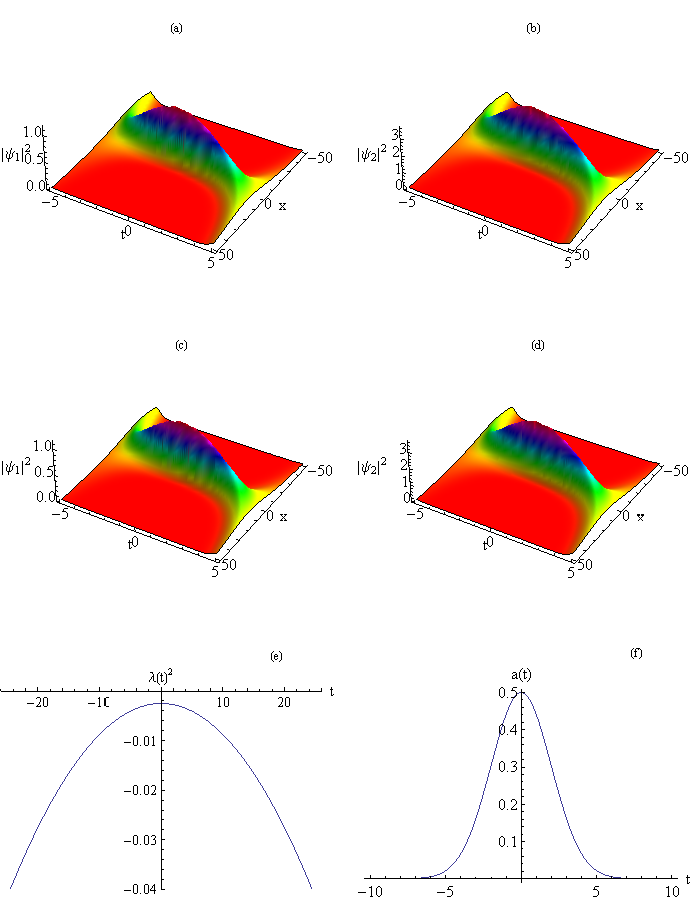}
\caption{{\bf Upper panel (a)-(b)}: Density (Analytical) profile
of the condensates by switching ON the time dependence in the
expulsive trap  for $\Gamma(t)$=-0.25 t and $a(t)=0.5 e^{-0.125
t^2}$; {\bf Middle Panel (c)-(d)}: Numerically simulated density
profile for $\Gamma(t)$=-0.25 t and $a(t)=0.5 e^{-0.125 t^2}$;
{\bf Lower panel (e)-(f)}: Transient trap and Interaction
strength} \label{fig.ana4}
\end{center}
\end{figure}

\begin{figure}
\begin{center}
\includegraphics[scale=0.25]{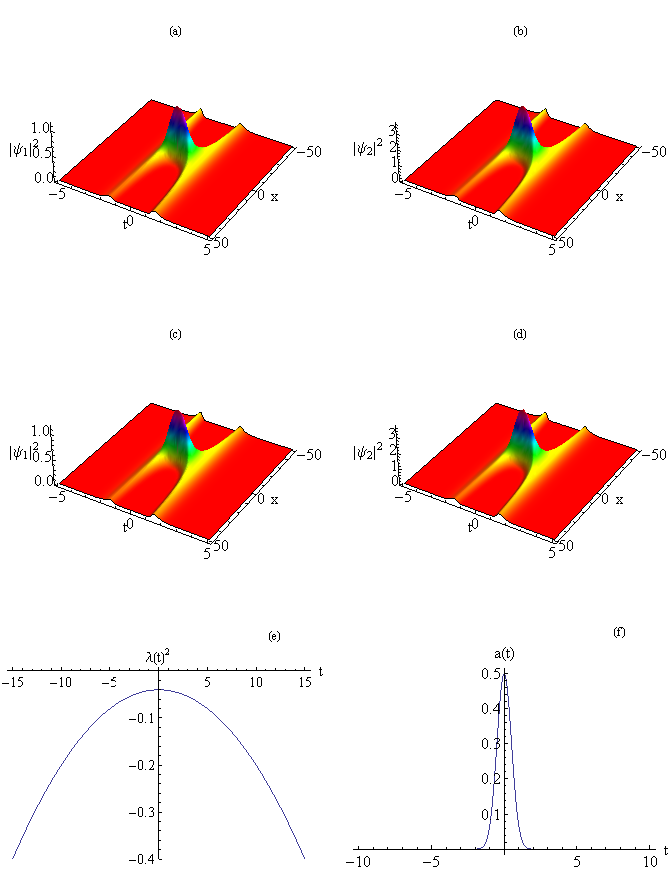}
\caption{{\bf Upper panel (a)-(b)}:Density (Analytical) profile of
the condensates in the time dependent expulsive trap for
$\Gamma(t)$=-5 t and $a(t)$=$0.5 e^{-2.5 t^2}$; {\bf Middle Panel
(c)-(d)}: Numerically simulated density profile for $\Gamma(t)$=-5
t and $a(t)$=$0.5 e^{-2.5 t^2}$; {\bf Lower panel (e)-(f)}:
Transient trap and Interaction strength}\label{fig.ana5}
\end{center}
\end{figure}

To enhance the stability of the condensates, we now switch on the
time dependence in the harmonic trap keeping it expulsive (shown
in Fig. \ref{fig.ana4} (e)) of the form $\Gamma(t)= -0.25 t$ and
manipulate the scattering length through Feshbach resonance of the
form $a(t)=0.5 e^{-0.125 t^2}$ shown in Fig. \ref{fig.ana4} (f)
consistent with equations (\ref{trap}) and (\ref{integ}), the
corresponding density profile is shown in Figs.\ref{fig.ana4}
(a)-(b). This is confirmed by the numerical simulation of the
condensates shown in Figs. \ref{fig.ana4} (c) and \ref{fig.ana4}
(d). When we make a 20 fold increase in the expulsive trap
frequency (shown in Fig. \ref{fig.ana5} (e)) and accordingly
employ Feshbach resonance to choose $a(t)=0.5 e^{-2.5 t^2}$ (shown
in Fig. \ref{fig.ana5} (f)) consistent with equations (\ref{trap})
and (\ref{integ}), the corresponding density profile shown in
Figs. \ref{fig.ana5}(a)-(b) exactly match with the numerically
simulated condensates  in Figs. \ref{fig.ana5} (c) and
\ref{fig.ana5} (d).

\begin{figure}
\begin{center}
\includegraphics[scale=0.25]{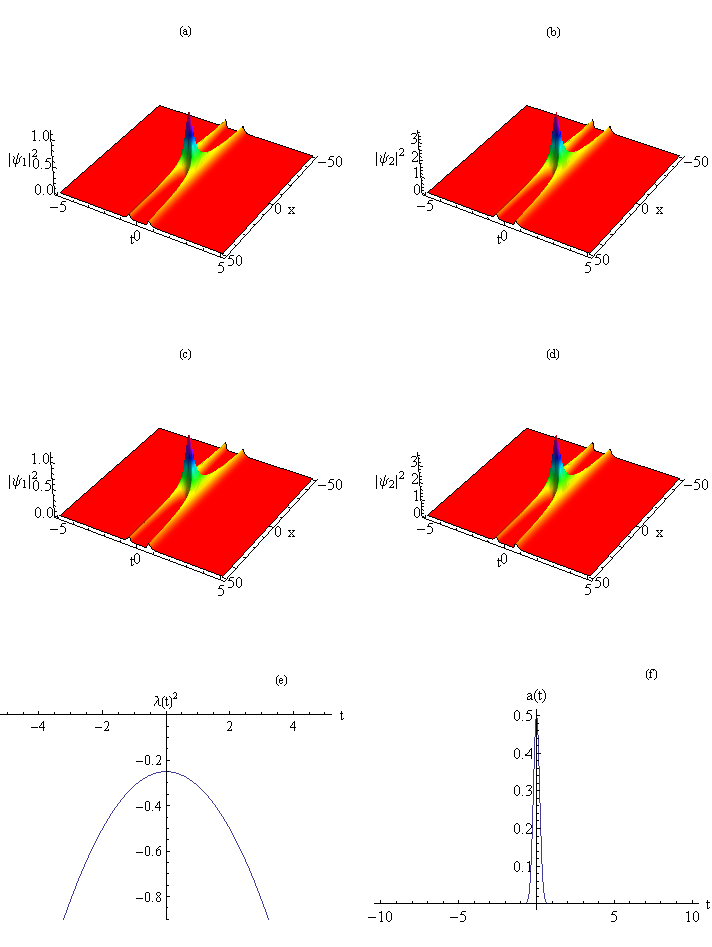}
\caption{{\bf Upper panel (a)-(b)}: Density (Analytical) profile
of the condensates in the time dependent expulsive trap for
$\Gamma(t)$=-25 t and $a(t)=0.5 e^{-12.5 t^2}$; {\bf Middle Panel
(c)-(d)}: Numerically simulated density profile for
$\Gamma(t)$=-25 t and $a(t)=0.5 e^{-12.5 t^2}$; {\bf Lower panel
(e)-(f)}: Transient trap and Interaction strength
}\label{fig.ana6}
\end{center}
\end{figure}

When we further enhance the time dependent expulsive trap (shown
in Fig. \ref{fig.ana6} (e)) frequency $\Gamma(t)$ by 100 times the
original strength  and manipulate scattering length (shown in Fig.
\ref{fig.ana6} (f)) through Feshbach resonance (in accordance with
equations (\ref{trap} and \ref{integ})), one never sees an abrupt
increase in the density of the condensates as shown in Figs.
\ref{fig.ana6} (a)-(b). Thus, we observe that any further increase
of the expulsive trap frequency $\Gamma(t)$ and manipulation of
scattering length through Feshbach resonance does not
significantly increase the density of the condensates eventhough
the attractive interaction strength increases rapidly. In other
words, the condensates in the time dependent expulsive harmonic
trap continue to remain stable for a reasonably large interval of
time even for large attractive interactions. Again, analytical
results synchronize with numerical simulations shown in Figs.
\ref{fig.ana6}(c)-(d).

Thus, we observe that the vector BECs in a time dependent
expulsive trap are more long lived and the lifetime can be
enhanced by Feshbach resonance management compared to the
condensates in a time independent expulsive harmonic trap. We also
wish to point out that we are able to stabilize the condensates in
an expulsive time dependent harmonic trap for attractive
interactions, where the condensates usually get compressed and
collapse subsequently. It is also pretty obvious that the scalar
counterpart of the condensates stabilized through Feshbach
resonance (shown in Figs. \ref{fig.ana4}-\ref{fig.ana6}) will
collapse immediately during time evolution.


\section{Stabilization of bright solitons in weakly coupled BECs}

\begin{center}
\textbf{A.   Impact of weak time dependent Rabi coupling}
\end{center}

We also emphasize that the enhancement in the lifetime of the two
component BECs arises by virtue of intraspecies and interspecies
interaction. It should be mentioned that one can make two
component BECs more longlived by the addition of weak time
dependent or space dependent coupling forces.

\begin{figure}
\begin{center}
\includegraphics[width=0.4\linewidth]{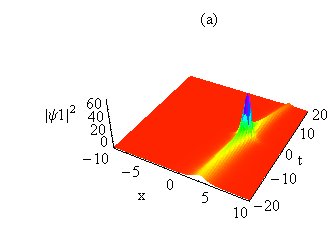}
\includegraphics[width=0.4\linewidth]{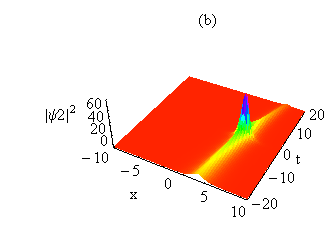}
\caption{Density  of the condensates for $a(t)=0.5t$,
$\varepsilon_1^{(1)}=0.3$ and $\Gamma(t)=0.1t$ X $10^{-2}$ without
coupling.}\label{psvf1}
\end{center}
\end{figure}
\begin{figure}
\begin{center}
\includegraphics[width=0.4\linewidth]{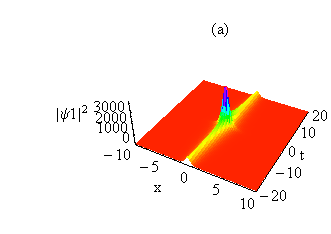}
\includegraphics[width=0.4\linewidth]{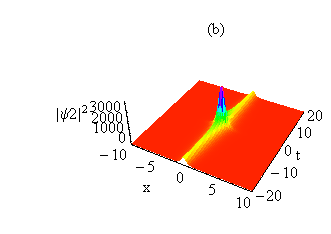}
\caption{Impact of weak coupling on the condensates with $\nu (t)
=  0.1 t $ with the other parameters as in
Fig.\ref{psvf1}}\label{psvf2}
\end{center}
\end{figure}
We now consider a spinor BEC composed of two hyperfine states, say
of the $ |F= 1, m_{f}=-1>$ and $ |F= 1, m_{f}=1>$ states of
$^{87}$Rb atoms \cite{rbatom} confined at different vertical
positions by parabolic traps and coupled by a time dependent
coupling field.

We assume the condensate to be quasi-one-dimensional
(cigar-shaped). Then, in the meanfield approximation, the system
is described by the coupled GP equation \cite{pethick}

\begin{eqnarray}
i\psi_{1t}+\psi_{1xx}&+&2a(t)(|\psi_1|^2
+|\psi_2|^2)\psi_1+\lambda(t)^2 x^2 \psi_1\notag\\
&+& i G(t)\psi_1+\nu(t)\psi_2= 0,\\
i\psi_{2t}+\psi_{2xx}&+&2a(t)(|\psi_1|^2
+|\psi_2|^2)\psi_2+\lambda(t)^2 x^2 \psi_2\notag\\
&+& i G(t)\psi_2 + \nu(t)\psi_1= 0,\label{lcgp}
\end{eqnarray}

In the above equation, $\nu(t)$ denotes the coupling between the
two condensates while $G(t)$ accounts for the feeding (loss/gain)
of the condensates from the thermal cloud. Infact, the impact of
adding a linear coupling to the Manakov model was already
demonstrated \cite{ref15} and the resulting dynamical system was
shown be integrable. The linearly coupled GP equations
(\ref{lcgp}) have also been investigated and the concept of domain
walls \cite{ref16} and symmetry breaking of solitons \cite{ref17}
was explored. Equations (\ref{lcgp})  have also been investigated
for $\nu(t)=0$ \cite{rajen09,twosol.vr.2010} and the dynamics of
vector BECs has been analyzed.
\begin{figure}
\begin{center}
\includegraphics[width=0.4\linewidth]{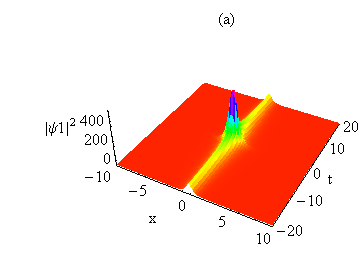}
\includegraphics[width=0.4\linewidth]{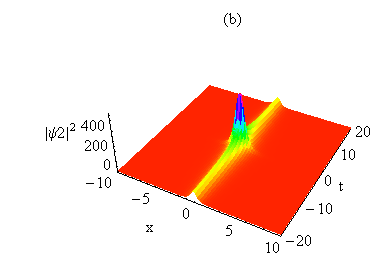}
\caption{Stabilization of the condensates  by tuning the
scattering length for  $a(t)= 0.5$ X $1/2 t$ with the other
parameters as in  Fig. \ref{psvf2}}\label{psvf3}
\end{center}
\end{figure}
The explicit forms of one soliton solution of equations
(\ref{lcgp}) can be written as \cite{ref37}
\begin{eqnarray}
\psi_{1}^(1) = \frac{2}{\sqrt a(t)}\varepsilon_1^{(1)}
\beta_{1}(t) sech(\theta_1)e^{i(-\xi_1 + \Gamma(t) \frac{x^2}{2})},\label{lcsol1}\\
\psi_2^(1) = \frac{2}{\sqrt a(t)} \varepsilon_2^{(1)} \beta_{1}(t)
sech(\theta_1)e^{i(-\xi_1 + \Gamma(t)
\frac{x^2}{2})},\label{lcsol2}
\end{eqnarray}
where
\begin{eqnarray}
\theta_1 &=& 8\int \alpha_1(t) \beta_1(t) dt+2 x \beta_1(t)-2 \delta_1,\\
 \xi_1 &=& 4\int(\alpha_1(t)^2-\beta_1(t)^2)dt+2x\alpha_1(t)-2\chi_1,
\end{eqnarray}
with $\alpha_1(t) = \alpha_{10}e^{-\int 2 \Gamma(t)^{2}dt}$,
$\beta_1(t)=\beta_{10}e^{-\int 2 \Gamma(t)^{2}dt}$ while
$\delta_1$ and $\chi_1$ are arbitrary parameters with
$\varepsilon_{1,2}$ as coupling parameters. Thus, it is obvious
from equation (\ref{lcsol1}) and (\ref{lcsol2}) that the amplitude
of the bright solitons depends on the temporal scattering length
$a(t)$ and trap frequency $\Gamma(t)$. ($\beta_{1} (t)$ varies
exponentially with $\Gamma(t)$). The fact that the trap frequency
$\Gamma(t)$ varies exponentially with the time dependent coupling
coefficient $\nu(t)$ (see  equation (9) in Ref. \cite{ref37})
indicates that the density of the bright solitons or the
condensates could build up in a very shot span of time. This means
that the dynamical system could enter into the domain of
instability very quickly.
\begin{figure}
\begin{center}
\includegraphics[width=0.45\linewidth]{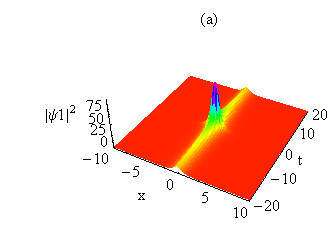}
\includegraphics[width=0.45\linewidth]{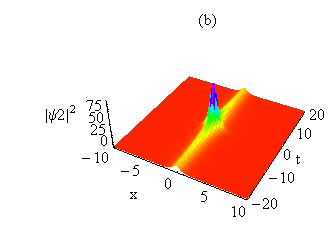}
\caption{Stabilization of the condensates by finetuning the
coupling coefficient for $\nu (t) = 0.1 t / 5 $ with the other
parameters as in  Fig. \ref{psvf2}}\label{psvf4}
\end{center}
\end{figure}
\begin{figure}
\begin{center}
\includegraphics[width=0.45\linewidth]{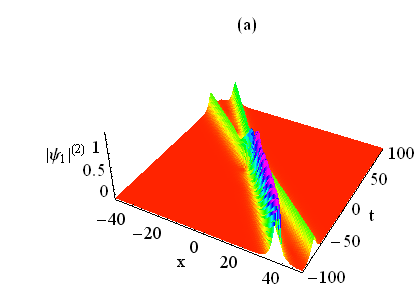}
\includegraphics[width=0.45\linewidth]{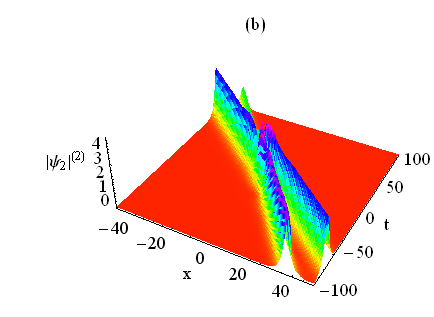}
\caption{Intramodal Inelastic collision of the condensates without
coupling for $a(t)=5, \Gamma(t)=0.1t$ X $10^{-2}$,
$\varepsilon_1^{(1)}=0.89i$ and
$\varepsilon_1^{(2)}=0.6$}\label{psvf5}
\end{center}
\end{figure}

The profile of the density of the condensates without the linear
Rabi coupling is shown in  Figs. \ref{psvf1} (a) and (b). When the
two condensates are coupled together, one observes an abrupt
increase in the density of the condensates as shown in Figs.
\ref{psvf2} (a) and (b). This suggests that coupling the two
condensates leads to an instability in the dynamical system.
However, this instability can be overcome either by changing the
temporal scattering length through Feshbach Resonance as shown in
Figs. \ref{psvf3} (a) and (b) or by finetuning the time dependent
coupling co-efficient as shown in  Figs. \ref{psvf4} (a) and (b).

\begin{figure}
\begin{center}
\includegraphics[width=0.45\linewidth]{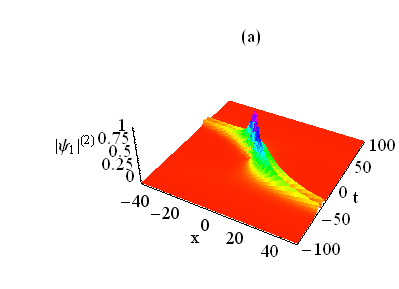}
\includegraphics[width=0.45\linewidth]{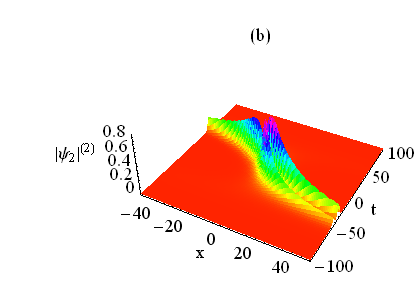}
\caption{Impact of coupling on the collision of condensates for
$\nu(t)=0.01 t$ with the other parameters as in Fig.
\ref{psvf5}}\label{psvf6}
\end{center}
\end{figure}
\begin{figure}
\begin{center}
\includegraphics[width=0.4\linewidth]{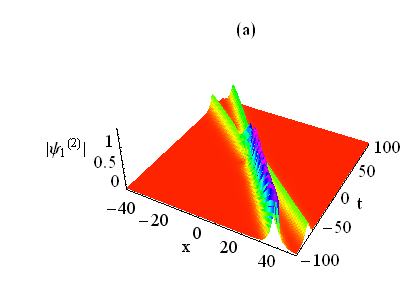}
\includegraphics[width=0.4\linewidth]{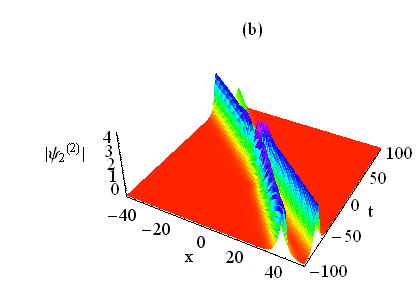}
\caption{Retrieval of intramodal inelastic interaction of the
condensates for $\nu(t)=0.01 t$X$10^{-2}$ with the other
parameters as in  Fig. \ref{psvf6}}\label{psvf7}
\end{center}
\end{figure}

The gauge transformation approach \cite{llc} can be easily
extended to generate multisoliton solutions. From the collisional
dynamics of the condensates, one observes an intramodal inelastic
interaction of bright solitons in the  absence of weak coupling as
shown in Figs. \ref{psvf5}(a) and (b). The addition of a weak time
dependent coupling disrupts the intramodal inelastic collision of
the condensates  as shown in Figs. \ref{psvf6}(a) and (b).
However, one can suitably finetune the coupling coefficient
$\nu(t)$ to retrieve the intramodal inelastic collision of the
condensates as shown in Figs. \ref{psvf7} (a) and (b).
\\

\begin{center}
\textbf{B.  Influence of weak spatially dependent coupling}
\end{center}
To understand the effect of weak spatially dependent coupling, we
now consider cigar-shaped (quasi-one-dimensional) BECs composed of
two hyperfine states of $^{87}$Rb atoms \cite{ref39} confined by
the parabolic trapping potential subject to a spatially dependent
force $\nu(x)$. The system is now described by the following
coupled Gross-Pitaevskii (GP) equation \cite{pethick} within the
framework of the mean-field description as
\begin{eqnarray}
i\psi_{1t}+\psi_{1xx}&+&2a(t)(|\psi_1|^2+|\psi_2|^2)\psi_1+
\lambda(t)^2 x^2 \psi_1\notag\\&+&i G(x,t)\psi_1
+\nu(x)\psi_2=0,\label{lcgps1}
\end{eqnarray}
\begin{eqnarray}
i\psi_{2t}+\psi_{2xx}&+&2a(t)(|\psi_1|^2+|\psi_2|^2)\psi_2+
\lambda(t)^2 x^2 \psi_2\notag\\&+&i G(x,t)\psi_2
+\nu(x)\psi_1=0,\label{lcgps2}
\end{eqnarray}
In the above equations, $\nu (x)$ denotes the weak spatial
coupling between the two components while $G(x,t)$ accounts for
the feeding which can be carried out by means of a reservoir
filled with a large amount of the condensate to which the trap is
weakly coupled \cite{ref41}. In fact, effects generated by adding
the linear coupling to the Manakov model were already studied
\cite{ref 15} and the resulting system was found to be integrable.
Equations (\ref{lcgps1}) and (\ref{lcgps2}) have been investigated
for $\nu =0$ \cite{rajen09,twosol.vr.2010}, and the ensuing
dynamics of the vectorial BECs has been analyzed. When the
trapping potential and (weak) spatial coupling between the two
condensates are neglected $(\lambda (t)=\nu (x)=0)$ and the
scattering length $a(t)$ becomes a constant (neglecting gain/loss
$G(x,t)$) then, the dynamical system becomes the celebrated
Manakov's model \cite{ref45}.

\begin{figure}
\includegraphics[scale=0.25]{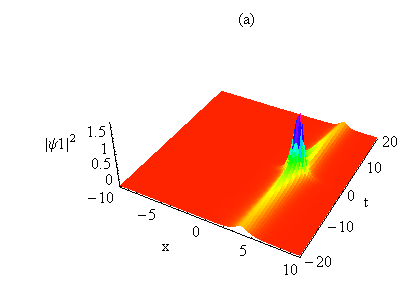}
\includegraphics[scale=0.25]{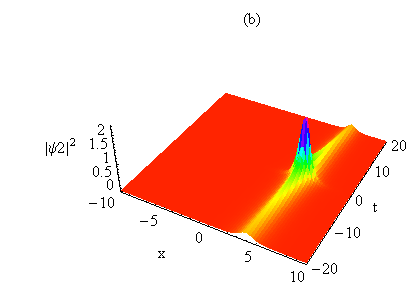}
\caption{The density of the condensates for $a(t)=0.5t$, $\protect%
\varepsilon _{1}^{(1)}=0.3$ and $\protect\Gamma(t)= \int 0.005 t $
dt in the absence of the coupling.}\label{rrp1}
\end{figure}

\begin{figure}
\includegraphics[scale=0.25]{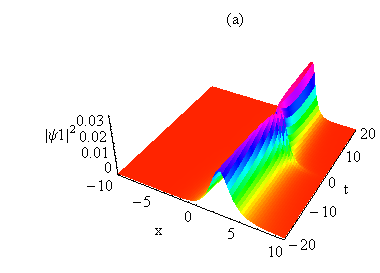}
\includegraphics[scale=0.25]{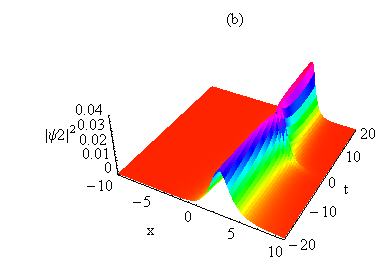}
\caption{The impact of weak linear coupling $\protect\nu (x)=0.5x$
on the condensates with the other parameters as in Fig.
\ref{rrp1}}\label{rrp2}
\end{figure}

The explicit forms of bright soliton solution of equations
(\ref{lcgps1}) and (\ref{lcgps2}) can be written as
\cite{rrpmypaper}
\begin{equation}
\psi _{1}^{(1)} ={\frac{2}{\sqrt{a}(t)}}\varepsilon
_{1}^{(1)}\beta _{1}(t) \mathrm{sech}(\theta _{1})e^{i(-\xi _{1}+
\Gamma (t)\int x dx)},\label{lcsolspatial1}
\end{equation}
\begin{equation}
\psi _{2}^{(1)} ={\frac{2}{\sqrt{a}(t)}}\varepsilon
_{2}^{(1)}\beta _{1}(t) \mathrm{sech}(\theta _{1})e^{i(-\xi _{1}+
\Gamma (t)\int x dx)},\label{lcsolspatial2}
\end{equation}%
where
\begin{eqnarray}
\theta_{1}&\equiv &2\int \beta_{1}dx+\int(8\alpha_{1}\beta
_{1}-4\alpha_{1})dt-2\delta_{1}, \\
\xi_{1}&\equiv &-2\int \alpha_{1}dx-4\int(\alpha_{1}+4i\beta
_{1}^{2}-4\beta_{1})dt-2\chi_{1},
\end{eqnarray}
with $\alpha_{1}(t)=\alpha_{10}e^{-2\gamma^{\prime}(t)}$, $\beta
_{1}(t)=\beta _{10}e^{-2\gamma ^{\prime }(t)}$, where $\delta _{1}$ and $%
\chi _{1}$ are arbitrary parameters and $\varepsilon _{1,2}$ are
coupling parameters.

\begin{figure}
\begin{center}
\includegraphics[scale=0.3]{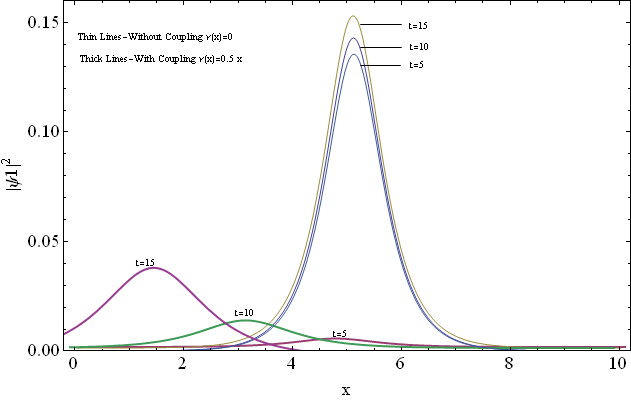}\caption{The
evolution of bright solitons (given by equation
(\ref{lcsolspatial1})) in the presence and absence of the linear
coupling in a confining trap for $\protect\Gamma (t)=\int 0.005 t$
dt.}\label{rrp3}
\end{center}
\end{figure}

Thus, it is obvious from equations (\ref{lcsolspatial1}) and
(\ref{lcsolspatial2}) that the amplitude of the bright solitons
depends on the time-modulated scattering length $a(t)$ and $\beta
_{1}{(t)}$, which implicitly depends on $\nu (x)$  (by virtue of
equation (7) in Ref. \cite{rrpmypaper}). The density profile of
the condensates shown in Fig. \ref{rrp1} (a), (b) and Fig.
\ref{rrp2} (a), (b) and the time evolution of the bright solitons
shown in Fig. \ref{rrp3} demonstrate that, in the absence of
spatial coupling, the density of the condensates grows with time
while they remain localized around the same point. The
introduction of the linear coupling not only stretches the wave
packet, but also shifts the center of the localization of the wave
packet as shown in Fig. \ref{rrp3}.
\begin{figure}
\begin{center}
\includegraphics[scale=0.325]{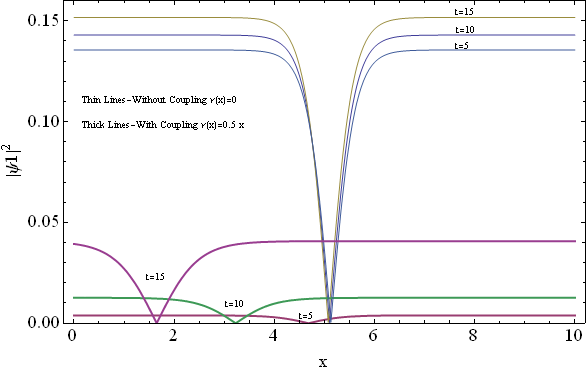}\caption{The
evolution of a dark soliton (given by equation
(\ref{lcsolspatialdark1})) in the presence and absence of the
spatially-dependent linear coupling in a confining trap for
$\protect\Gamma (t)=\int 0.005 t$ dt.}\label{rrp4}
\end{center}
\end{figure}

The explicit forms of the dark-soliton solution can be written as
\begin{equation}
\psi _{1}^{(1)} = \frac{2}{\sqrt{a(t)}}\varepsilon
_{1}^{(1)}\beta_{1}(t)\mathrm{tanh}(\theta_{1})e^{i(-\xi_{1}+
\Gamma (x,t)\int x dx)}\label{lcsolspatialdark1}\\
\end{equation}
\begin{equation}
\psi_{2}^{(1)}= {\frac{2}{\sqrt{a(t)}}\varepsilon _{2}^{(1)}\beta
_{1}(t)\mathrm{tanh}(\theta_{1})e^{i(-\xi _{1}+ \Gamma (x,t)\int x
dx})}\label{lcsolspatialdark2}
\end{equation}
The time evolution of the dark solitons indicates that one
observes a similar impact of the linear coupling (a shift of the
center of the localization as shown in Fig. \ref{rrp4} and
stretching of the matter wave packet) in the dark solitons as
well.
\begin{figure}
\begin{center}
\includegraphics[scale=0.2]{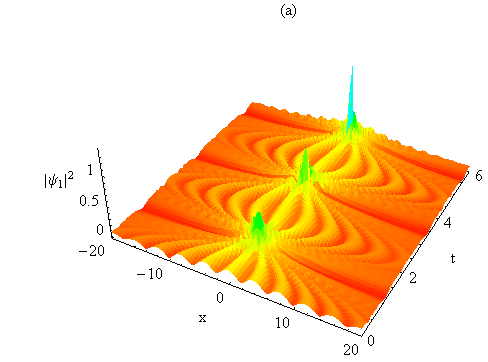}
\includegraphics[scale=0.2]{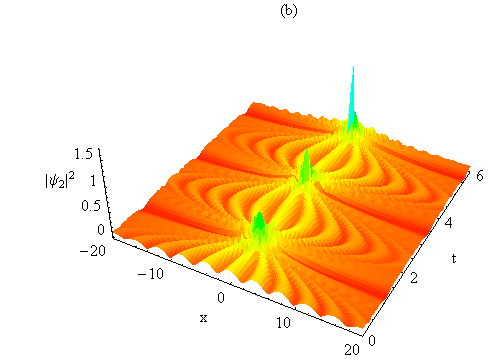}
\caption{Density profiles of bright-bright soliton interaction for
the parametric choice $a(t)=-\cos
(\protect\sqrt{2}t),\protect\Gamma(t)=\int 0.001 t$ dt,
$\protect\nu (x)=\left[ \log (-\cos \left(
2x\right) )\right] ^{2}$, $\protect\varepsilon _{1}^{(1)}=0.89i,$ and $%
\protect\varepsilon _{1}^{(2)}=0.6$ in an expulsive
trap.}\label{rrp5}
\end{center}
\end{figure}

\begin{figure}
\begin{center}
\includegraphics[scale=0.25]{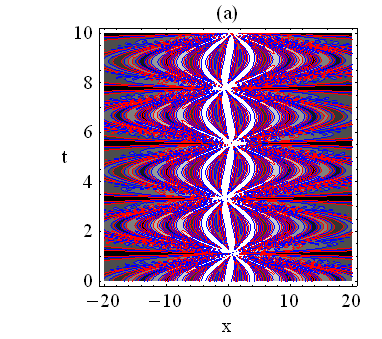}
\includegraphics[scale=0.25]{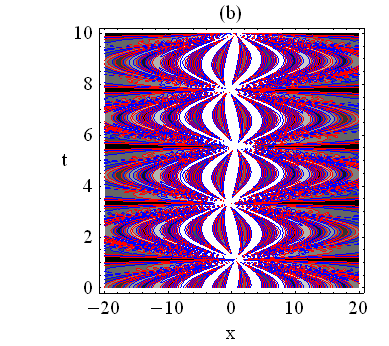}
\caption{Contour plots for the bright-bright soliton interaction
shown in Fig. \ref{rrp5}.}\label{rrp6}
\end{center}
\end{figure}

\begin{figure}
\begin{center}
\includegraphics[scale=0.25]{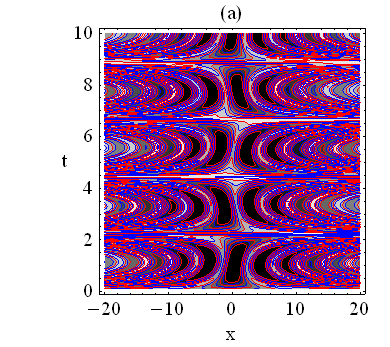}
\includegraphics[scale=0.25]{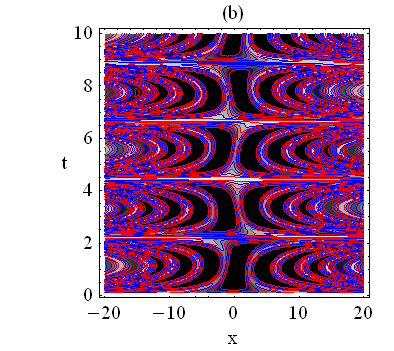}
\caption{The interference pattern arising from the dark-dark
soliton interaction.}\label{rrp7}
\end{center}
\end{figure}

Another interesting observation is that the addition of the
spatially-dependent linear coupling does not contribute to the
growth of the condensates unlike the time-dependent coupling
\cite{ref37}. Instead, the spatial coupling merely stretches the
wave packet. This means that one can easily stabilize the
vectorial BECs by the introduction of the spatially-dependent
linear coupling. One can easily extend gauge transformation
approach to generate multiple dark solitons.

The shift of the center of the matter-wave packet, which is
induced by the spatially-dependent coupling, can be suitably
exploited to generate matter-wave interference pattern which was
addressed theoretically \cite{ref51}. The density profile for
bright-bright soliton interaction is shown in Figs. \ref{rrp5} (a)
and (b). From the density profile, one observes that the matter
wave density reaches the maximum value periodically and the
central fringe is always bright in every interference pattern.
This means that the bright - bright solitons constructively
interfere at the centre giving rise to maximum intensity. One
observes alternate bright and dark fringes in the pattern. The
contour plot shown in Fig. \ref{rrp6} confirms this observation.

The contour plot corresponding to the density profile of dark-dark
soliton interaction is shown in Fig. \ref{rrp7}. Interference
pattern shown in Fig. \ref{rrp7} indicates that the central fringe
is always dark indicating that the dark-dark solitons
destructively interfere at the centre. Again, one notices
alternate dark and bright fringes.

\begin{figure}
\begin{center}
\includegraphics[scale=0.3]{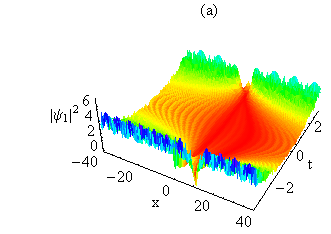}
\includegraphics[scale=0.3]{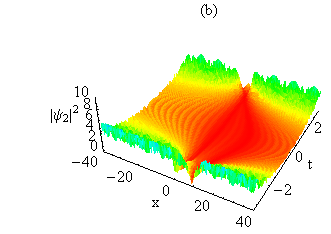}
\caption{Density profiles for the dark-bright soliton interaction
in a transient trap for the parametric choice $\protect\nu (x)=\left[ \tan (0.03x)%
\right] ^{2}+0.0075$, $a(t)=0.001 t $, $\protect\Gamma (t)=-%
\protect\int 0.05te^{-0.005t}dt$, $\protect\varepsilon
_{1}^{(1)}=0.89i$, and $\protect\varepsilon
_{1}^{(2)}=0.6$.}\label{rrp8}
\end{center}
\end{figure}

\begin{figure}
\begin{center}
\includegraphics[scale=0.35]{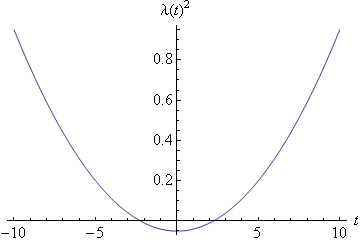}\caption{The transient trap for $\protect\Gamma (t)=-\protect\int %
0.05te^{-0.005t}dt$.}\label{rrp9}
\end{center}
\end{figure}

\begin{figure}
\begin{center}
\includegraphics[scale=0.35]{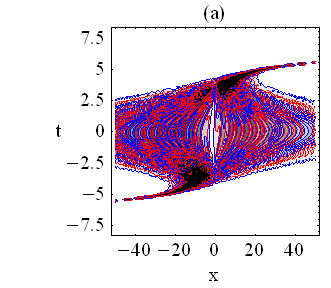}
\includegraphics[scale=0.35]{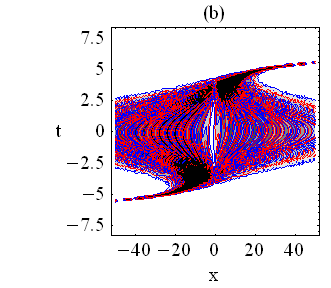}
\caption{Contour plot of the the dark-bright soliton interaction
shown in Fig. \ref{rrp8}.}\label{rrp10}
\end{center}
\end{figure}

Choosing a transient trap of the form shown in Fig. \ref{rrp9},
the density profile of dark-bright interaction is shown in Fig.
\ref{rrp8}. From the figures, we understand that when the solitons
make a transition from the expulsive region to the confining
region, the dark and bright solitons interfere with each other in
the confining domain. When the trap gets back to the expulsive
domain again, the interference pattern disappears while the
density of the condensates gets compressed. The above interference
pattern is quite identical to the one observed in Ref.
\cite{rames09pla}. From the interference pattern shown in Fig.
\ref{rrp10}, one observes that the central fringe  is always
bright indicating the maximum intensity. This occurs because  the
amplitude of bright solitons predominates over that of the dark
solitons. It must be mentioned that the spacing between the bright
(or dark) fringes in the interference pattern represented by Figs.
\ref{rrp6} and \ref{rrp7} gives a measure of the coherence of
matter waves and is a characteristic feature of the atomic system
under investigation. The spacing between the bright (or dark)
fringes in the interference pattern shown in Fig. \ref{rrp10}
observed in the confining trap has narrowed down as the
condensates dwell in the confining trap for a very short interval
of time.

\section{Future Directions}

In the present review, we have analytically proved that vector
BECs are longlived compared to scalar BECs. Eventhough this paper
has comprehensively reviewed the recent developments from the
perspective of integrabily in scalar and vector BECs in a time
independent/ dependent harmonic trap, there are several unexplored
territories in the domain of ultra cold atoms. The identification
of an integrable model for a two component BEC (either for a BEC
comprising of hyperfine states of the same atom or two different
atoms) characterized by three distinct scattering lengths, namely
two distinct intraspecies scattering lengths and an interspecies
scattering length has eluded our observation so far. Eventhough
weak coupling of the condensates generates high density matter
waves and one can somehow stabilize them, the question of
nonlinearly (strongly) coupling the condensates remains
unanswered. These investigations will certainly enable us to
penetrate deep into the dynamics of ultra cold atoms with
precision. Investigations along these directions are in progress
and the results will be published later.

\emph{Acknowledgements.}  RR wishes to acknowledge the financial
assistance received from DST (Ref.No:SR/S2/HEP-26/2012 dated
16.10.2012), UGC (Ref.No:F.No 40-420/2011(SR) dated 4.July.2011),
DAE-NBHM (Ref.No: NBHM/R.P.16/2014/Fresh dated 22.10.2014) and
CSIR (Ref.No: No.03(1323)/14/EMR-II dated 03.11.2014). RR and PSV
wish to acknowledge the contribution of Dr. V. Ramesh Kumar in
sharing the results of his doctoral thesis. PSV wishes to thank
UGC for a Senior Research Fellowship.

\end{document}